\documentclass[prb,aps,reprint,twocolumn]{revtex4-1}

\usepackage{graphicx}
\usepackage{amsmath}
\usepackage{amsfonts}
\usepackage{amssymb}

\usepackage{epstopdf}
\usepackage[usenames,dvipsnames]{xcolor}

\newcommand{\mean}[1]{\langle #1 \rangle}
\newcommand{\boldgreek}[1]{\ensuremath{\mbox{\boldmath$#1$}}}

\newcommand{\ddx}[1]{#1_{,x}}

\newcommand{\bee}{b}

\newcommand{\rhoG}{\rho_{\rm G}}
\newcommand{\rhoS}{\rho_{\rm S}}
\newcommand{\cT}{\ensuremath{c_{\rm T}}}
\newcommand{\cL}{\ensuremath{c_{\rm L}}}

\renewcommand{\Re}{{\rm Re}\;}
\renewcommand{\Im}{{\rm Im}\;}
\renewcommand{\imath}{i}

\newcommand{\FT}[1]{\hat{#1}}

\begin{document}

	\title{Nonlinear Damping in Graphene Resonators}
	\author{Alexander Croy}
	\email{alexander.croy@chalmers.se}
	\author{Daniel Midtvedt}
	\author{Andreas Isacsson}
	\author{Jari M. Kinaret}
	
	\affiliation{Department of Applied Physics, Chalmers University of Technology, 
					S-412 96 G\"oteborg, Sweden}

	\date{\today}

	\begin{abstract}
			Based on a continuum mechanical model for single-layer graphene we propose and analyze a microscopic mechanism for dissipation in nanoelectromechanical graphene resonators. We find that coupling between flexural modes and in-plane phonons leads to linear and nonlinear damping of out-of-plane vibrations. By tuning external parameters such as bias and ac voltages, one can cross over from a linear to a nonlinear-damping dominated regime. We discuss the behavior of the effective quality factor in this context.
	\end{abstract}

	\maketitle

\section{Introduction}
	Advances in fabrication and detection techniques have enabled a wide range of
	experimental realizations of carbon-based nanoelectromechanical (NEM) resonators
	\cite{buza+07,erle+08,chro+09,eimo+11}.	
	However, to optimize their operation, an increased understanding of dissipation 
	mechanisms is needed. For NEM resonators in general, several
	processes leading to linear damping (LD) have been investigated 
	\cite{liro00,crli01,wil08,rebl+09}.
	Specifically for graphene, at high temperatures, ohmic losses in
	the metallic gate and the graphene sheet have been argued to
	limit the quality factor \cite{segu+07}.
	Recently, the focus has shifted to study quantum aspects of mechanical 
	motion \cite{ocho+10,tedo+11}, such as mechanical cat states \cite{voki+12}, 
	which require a more 
	detailed understanding of dissipation and decoherence mechanisms.

	Since graphene-based resonators exhibit nonlinear behavior,
	one can expect the damping also to be amplitude dependent 
	\cite{dykr84,licr08,zash+12}. 
	Nonlinear damping (NLD) was reported
	in recent experiments on graphene and carbon nanotube resonators \cite{eimo+11}.
	However, little is known about the underlying physical mechanism,
	and typically phenomenological models are employed \cite{dykr84,licr08,zash+12}. 
	In these models, the resonator is coupled to a bath of harmonic oscillators. For
	couplings that depend quadratically on the resonator amplitude, it is known that
	NLD emerges \cite{zw73,lise81,dykr84}.
	
	For carbon-based resonators such a coupling naturally arises if the strain
	couples linearly to the degrees of freedom of some subsystem, which can be regarded
	as a bath. Two examples are the interaction between phonons and electrons\cite{cagu+09,vogu+09} and the coupling of mechanical modes. The relative importance
	of the two mechanisms is {\em a priori} not known and will also depend on the details
	of the experimental realization.
	
	In order to quantify the importance of the mechanical dissipation channel 
	for NLD, we analyze the coupling between flexural modes and in-plane phonons. We show that 
	it leads to a quadratic coupling and, consequently, 
   to both LD and NLD. Whether LD or NLD dominates
   is determined by the ratio of vibrational amplitude and static deflection. 
   We give an estimate for the expected crossover between LD and NLD, which can 
   be experimentally verified.

   \section{Model and Method}
\begin{figure}
		\centering
		\includegraphics[width=\columnwidth]{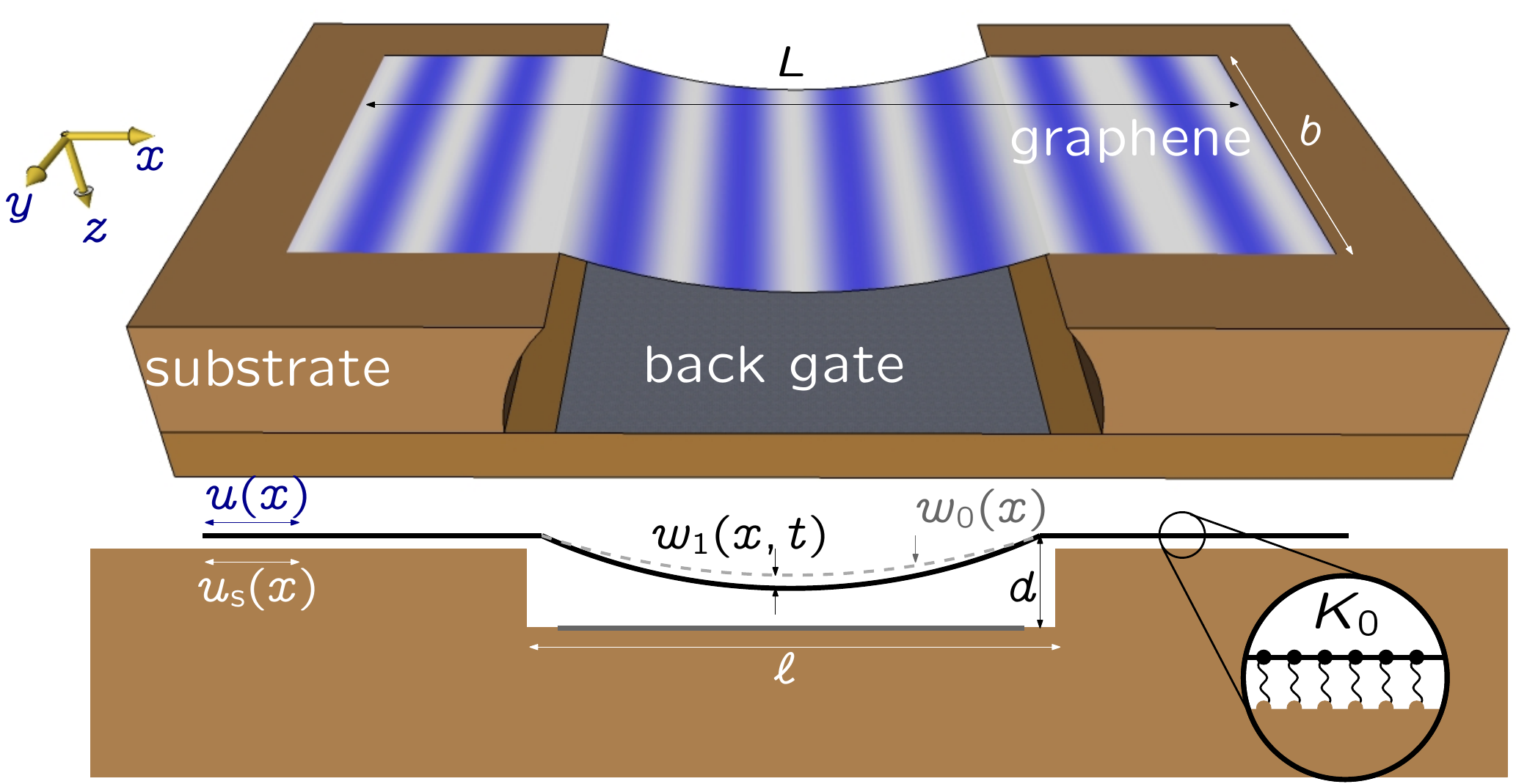}
		\caption{(Color online) Schematic view of a suspended graphene membrane 
		over a trench
		in an insulating substrate. A metallic gate is used for actuating 
		the resonator. In-plane phonons
		are created in the
		suspended region and dissipate energy as they propagate away.}
		\label{fig:setup}
	\end{figure}

   We consider 
	a graphene sheet of length $L$ and breadth $\bee$, suspended over a trench of
	width $\ell$ (cf.\ Fig.\ \ref{fig:setup}).	
	The van der Waals attraction between the graphene and the substrate 
   clamps down the sheet outside the suspended region \cite{sase+08,kobo+11,kuca+11}. 
	The trench is modeled by allowing the sheet to freely displace vertically in
	this region.
	Since out-of-plane 
	displacement is accompanied by in-plane stretching or compression, 
	flexural motion is converted into in-plane phonons in the suspended region.
	The clamping constrains the out-of-plane
	motion over the substrate, but still allows for small in-plane displacements. 
	Consequently, in-plane phonons created in the suspended region 
	transport energy away from this region. 
	In contrast to a phenomenological modeling approach we can relate dissipation
	to specific properties of the substrate and the graphene-substrate coupling. These properties
	can be obtained independently by theoretical or experimental means.
The dynamics of graphene NEM-resonators are well
described by the continuum theory of 2D-membranes \cite{nepi+88}. 
For a resonator made from a sheet lying in
the $xy$-plane, this theory is conveniently formulated in terms of the
in-plane
displacement fields $u(x,y),v(x,y)$ in the $x-$ and $y-$ directions,
respectively, and the displacement field in the $z-$direction,
$w(x,y)$. The equations of motion follow from the free energy ${\cal F}=\int
dx dy\, [{\cal F}_b + {\cal F}_s]$ where ${\cal F}_b=\frac{\kappa}{2}|\Delta w|^2$ is the free energy density
associated with pure bending and ${\cal F}_s=\frac{1}{2}\sum_{i,j}\sigma_{ij}\epsilon_{ij}$ is associated
with stretching of the membrane. The symmetric 2D strain and stress
tensors are here defined as
\begin{subequations}
\begin{multline}
\epsilon_{xx}=u_{,x}+w_{,x}^2/2,\quad
2\epsilon_{xy}=(u_{,y}+v_{,x})+w_{,x}w_{,y},\\
\epsilon_{yy}=v_{,y}+w_{,y}^2/2\;,
\end{multline}
and
\begin{multline}
\sigma_{xx}=(\lambda_{\rm G}+2\mu_{\rm G})\epsilon_{xx}+\lambda_{\rm G}\epsilon_{yy},\quad
\sigma_{xy}=2\mu_{\rm G}\epsilon_{xy}, \\
\sigma_{yy}=(\lambda_{\rm G}+2\mu_{\rm G})\epsilon_{yy}+\lambda_{\rm
 G}\epsilon_{xx}\;,
\end{multline}
\end{subequations}
respectively. Spatial derivatives are denoted by
subscripts, {\em i.e.}, $\ddx{u}=\partial u/\partial x$.
The expression for the free energy, which is similar to that for large
deflections of a plate \cite{lali86}, contains three
material parameters, the bending energy $\kappa\approx 1.1 - 1.6$~eV, and
the Lam\'e parameters, $\mu_{\rm G}\approx 146$~N/m and $\lambda_{\rm
 G}\approx 48$~N/m for graphene \cite{yabr+96,falo+07,atis+08,limi+12}.
To study qualitatively the effect of phonon radiation into the
supporting substrate, we assume for simplicity a quasi 1D situation
where variations in $y-$direction are disregarded. This would be valid
for a wide sheet where deviations from this assumption is confined to
the regions around the edges. In this case we have only the
displacement fields $u(x,t)$ and $w(x,t)$.  In any realistic
functioning device, there is some small amount of built in strain. In
practice, this implies that the energy contribution from the bending
energy is always negligible for the lowest lying flexural
modes \cite{atis+08}. Hence, to a good approximation we have for the quasi 1D
graphene resonator attached to a substrate the free energy density
\begin{multline}\label{eq:1dfreeen}
   \mathcal{F}(x,y) =
                     \frac{T_1}{2} \left(
                        \ddx{u}^2 + \ddx{u}\ddx{w}^2 + \frac{1}{4}
\ddx{w}^4
                    \right) \\
                    +\frac{1}{2} K(x) \left( u - u_{\rm S}\right)^2
                    + \mathcal{E}_{\rm ext}[w]\;,
\end{multline}
where we have defined $T_1 = \lambda_{\rm G}+2\mu_{\rm G}$. The
potential $\mathcal{E}_{\rm ext}[w]$ accounts for interactions used to
actuate the resonator. The second to last term couples the graphene
displacement to the substrate displacement $u_{\rm S}(x,y)$ in a
harmonic approximation \cite{swan99}, which largely allows us to
obtain an analytical description.

The function $K(x)$ restricts this coupling to the supported region, {\em i.e.}, $K(x)=K_0\Theta(|x|-\ell/2)$
with $\Theta$ being the Heaviside step function. The substrate is modeled as an elastic half-space and displacement at the surface, $\vec{s}(\vec{x},z=0,t) = (u_{\rm S}, v_{\rm S}, w_{\rm S})$, is given in terms of a response function \cite{lali86,pe01,mami76},
\begin{multline}\label{eq:sub_resp}
    s_{\mu}(\vec{x},z=0,\omega) = -\sum_{\nu}\int \frac{d^2 x'}{(2\pi)^2} R_{\mu \nu}(\vec{x}-\vec{x}',\omega) \\\times\sigma_{\nu z}(\vec{x}',\omega)\;.
\end{multline}
Consistent with the 1D model of the graphene sheet, only $\overline{u}_{\rm S}(x)\equiv \int^{\bee/2}_{-\bee/2} dy\, u_{\rm S}(x,y)$ is considered. 
Within the harmonic approximation, $\sigma_{xz}=K(x)\left(u-u_{S}\right)$.

The free energy \eqref{eq:1dfreeen} leads to a coupling between flexural
vibrations and in-plane motion via the coupling energy $\mathcal{E}_{\rm coup} = (T_1/2) \ddx{u}\ddx{w}^2$, which
is nonlinear in the flexural vibration amplitude. This coupling leads to NLD of the flexural vibrations \cite{zw73,lise81,dykr84,zash+12}.
 
\subsection{Equations of motion}\label{sec:eom}
The equations of motion for the out-of-plane and in-plane vibrations resulting from Eq.\ \eqref{eq:1dfreeen} are
\begin{subequations}\label{eq:eom}
\begin{align}
\rhoG \ddot{w} - \frac{T_1}{2} \frac{d}{dx} \left( 2 \ddx{u} \ddx{w} + \ddx{w}^3 \right)
    &= f_{\rm dc} + f_{\rm ac} \cos(\Omega t)\;, \label{eq:w_eom}\\
\rhoG \ddot u-  \frac{T_1}{2} \frac{d}{dx}\left(2 \ddx{u} + \ddx{w}^2\right)&=-K(x)\left(u-\overline{u}_{\rm S}/\bee \right), \label{eq:u_eom}
\end{align}
\end{subequations}
where $f_{\rm dc}(x)$ and $f_{\rm ac}(x) \cos(\Omega t)$ are the static and time dependent parts of the actuation force. Typically, electrostatic actuation is used, resulting from a time dependent back-gate voltage of the form $V_{\rm bg}(t)=V_{\rm dc}+V_{\rm ac}\cos(\Omega t)$ with $V_{\rm dc}\gg V_{\rm ac}$. To simplify the analysis, we assume the equilibrium stress field resulting from $f_{\rm dc}$ to be spatially uniform and equal to the tensile stress $T_0$ on the boundary \cite{lali86}. 
Generally, at a given back-gate bias voltage, the resonance frequency $\Omega_0(V_{\rm dc})$ depends
on initial stress and contains a shift due to electrostatic forces.
This so-called tuning behavior will be further discussed in Sec.\ \ref{sec:res_freq}.

Since Eq.\ \eqref{eq:u_eom} is linear in $u$, the influence of the environment can be accounted for
by a Green's function embedding technique. The solution,
\begin{equation}
	u(x,t) = \int dx' \int dt' G(x,x',t-t') \frac{c^2}{2} \frac{d}{dx'} {w_{,x'}}^2(x',t')\;,
\end{equation}
is given in terms of the in-plane response function $G$, which contains information about the attachment to the substrate via Eq.\ \eqref{eq:sub_resp}. The speed of sound in graphene is denoted by $c=\sqrt{T_1/\rhoG}$, where $\rho_{\rm G}$ is the mass density of graphene.

\subsection{Flexural mode dynamics}
Next, we consider the fundamental flexural mode and set $w(x,t) = q(t) \phi(x)$ for $|x| \leq \ell/2$ and zero otherwise. The mode shape $\phi$ is normalized to the length of the resonator. 
Upon projecting Eq.\ \eqref{eq:w_eom} onto the fundamental mode, an ordinary differential equation
for the vibration amplitude $q$ is obtained.
Further, moving to a rotating frame, we write $q(t)=\left[q_0 +\frac{1}{2}\left(q_1(t)e^{i\Omega t}+q_1^*(t)e^{-i\Omega t}\right)\right]$ and $\dot{q}(t)=\frac{i\Omega}{2}\left[q_1(t)e^{i\Omega t}-q_1^*(t)e^{-i\Omega t}\right]$. Inserting these expressions into
the equation of motion and performing the averaging yields an equation for the
slowly varying amplitude $q_1$ [\onlinecite{dykr84}], which contains memory terms related to linear
and non-linear damping. As the time-scales for flexural motion and in-plane phonons
are well separated ($\Omega_0\ll c/\ell$), the memory terms can be eliminated. 
This procedure corresponds to a Markov approximation \cite{dykr84}.
It is convenient to define new quantities
\begin{multline}\label{eq:ovrlp}
\FT{\chi}(\Omega) = \frac{c^2}{2}\int\limits_{-l/2}^{l/2}dx\int\limits_{-l/2}^{l/2}dx' \frac{d}{dx}\left[\ddx{\phi}^2  \FT{G}( x,x', -\Omega)\right] \\\times\frac{d}{dx'}\phi_{,x'}^2\;,
\end{multline}

where $\FT{G}( x,x', \omega) = (2\pi)^{-1}\int d\tau G(x,x',\tau) e^{\imath \omega \tau}$ is
the Fourier transform of the in-plane response function.

We obtain an equation of motion for the complex envelope function
\begin{multline}\label{eq:env_eom}
    m \dot{q_1} =
    \left[
        \imath m \left(\Omega_0-\Omega\right) q_1
        + \imath \frac{3}{8}\frac{\alpha}{\Omega_0} |q_1|^2 q_1 \right.\\\left.
        -\frac{1}{2} \gamma q_1
        -\frac{1}{8} \eta |q_1|^2 q_1 
        - \frac{\imath}{2\Omega_0} g
    \right]\;.
\end{multline}
For finite temperatures this equation has to be supplemented by noise forces, satisfying the fluctuation-dissipation relations. The thermally induced vibrations can lead to an additional broadening of
the response curves \cite{dykr84,basa+12}. In order to obtain a lower bound of LD and NLD we will work in the limit of zero temperature.
In Eq.\ \eqref{eq:env_eom}, the coefficients $m=\rhoG \ell \bee$, $\alpha$, $\gamma$ and $\eta$ denote the suspended mass, the Duffing elastic constant, linear and non-linear damping, respectively. They are given in terms of $\FT{\chi}$ as follows
\begin{subequations}\label{eq:env_coeff}
\begin{align}
    {\alpha} &=  
    {\alpha_0} - \frac{T_1 \bee}{2} \frac{4}{3}\Re \left(\FT{\chi}(0) + \frac{1}{2}\FT{\chi}(2\Omega)\right) \;,\\
    \gamma &=  -\frac{T_1 \bee}{2\Omega_0} q_0^2 \,4 \Im \FT{\chi}(\Omega) \;,\\
    \eta &=  -\frac{T_1 \bee}{2\Omega_0} \,2 \Im \FT{\chi}(2\Omega)\;.
\end{align}
\end{subequations}
Here, the bare Duffing constant is given by $\alpha_0 = (T_1 \bee/2) \int dx\, \ddx{\phi}(x)^4$.
The driving strength is $g = \int dx\, \phi(x) f_{\rm ac}(x)$. 
In accordance with our previous simplifications, we neglect the small polaronic shift of $\Omega_0$,
which is proportional to $\Re \FT{\chi}$, and an additional shift of $\alpha$ due to the
broken symmetry in the presence of static deflection.
Equation \eqref{eq:env_eom} is similar to the
equations used to model NLD in 
micromechanical resonators \cite{licr08,zash+12} and recent 
experiments on carbon-based resonators 
\cite{eimo+11}, the difference being the dependence of the
damping coefficients in Eq.\ \eqref{eq:env_coeff} on the driving frequency. 

In Eq.\ \eqref{eq:env_eom} the prevailing damping mechanism is determined by the ratio
\begin{equation}\label{eq:lin-nonlin}
	\tilde{\delta} \equiv \frac{\eta |q_1|^2}{4\gamma} 
		            \approx \frac{{\Im} \FT{\chi}(2\Omega)}{8 {\Im}\FT{\chi}(\Omega)}
		            			  \frac{|q^{\rm max}_1|^2}{q_0^2}\;.
\end{equation}
Here, $|q^{\rm max}_1|$ denotes the maximum amplitude of the response for a given driving strength.
Thus, $\tilde{\delta}$ is determined by the ratio of the overlap integrals defined in Eq.\ \eqref{eq:ovrlp}, which are purely geometrical quantities, and the ratio between the vibrational amplitude and the static deflection. For a small static deflection, it is therefore expected that NLD dominates the damping 
caused by phonon radiation.
Similarly, the dimensionless ratio  
\begin{equation}\label{eq:teta}
	\tilde{\eta} = \frac{\eta\; \Omega_0}{\alpha}
\end{equation}
measures the relative importance of the two nonlinearities in Eq.\ \eqref{eq:env_eom}
\cite{licr08}. 
For $\tilde{\eta}<\sqrt{3}$, the well-known bifurcation of the Duffing equation is present, while for $\tilde{\eta}>\sqrt{3}$ this bifurcation vanishes. The ratio $\tilde{\eta}$ is also a purely geometrical factor, apart from the weak dependence of $\Omega_0$ on the static deformation of the graphene. 

\subsection{Numerical method}\label{sec:num_method}
To compute the overlap integrals \eqref{eq:ovrlp} we first consider the Fourier transformed 
response of the substrate \eqref{eq:sub_resp} 
\begin{align}
\overline{u}_{\rm S}(x, \omega)  ={}& 
     -\int\limits^{L/2}_{-L/2} \frac{d x'}{(2\pi)^2} 
    	\int\limits^{\bee/2}_{-\bee/2} dy'\,
	   \int\limits^{\bee/2}_{-\bee/2} dy\, \notag\\
	   &	\quad\times R_{x x}(x-x',y-y',\omega) \sigma_{x z}(x',y',\omega) \notag\\
	   \approx{}&
	   -\int\limits^{L/2}_{-L/2} \frac{d x'}{(2\pi)^2} 
		\overline{R}_{xx}(x-x', \omega)\, \overline{\sigma}_{x z}(x',\omega)
	   \;. \label{eq:sub_resp_bar}
\end{align}
In the second step, in order to get a purely 1D response function, we have approximated the $y'$-dependence of $\sigma_{x z}(x',y')$ by the mean value $\frac{1}{\bee}\overline{\sigma}_{x z}$ and defined $\overline{R}_{xx}(x-x',\omega)\equiv \frac{1}{\bee}\int^{\bee/2}_{-\bee/2} dy'\,
\int^{\bee/2}_{-\bee/2} dy \, R_{x x}(x-x',y-y',\omega)$ \footnote{We found that $\overline{R}_{xx}(x-x',\omega)$ is well approximated by the integral $\int^{\bee/2}_{-\bee/2} dy\, R_{xx}(x-x',y,\omega)$.}.
The response function $R_{\mu \nu}$ for an elastic half-space
is known analytically \cite{lali86,pe01,mami76} and mainly depends on the longitudinal
and transversal sound velocities of the substrate (see Appendix \ref{sec:app_response}).

Evaluating Eq.\ \eqref{eq:sub_resp_bar} at discrete positions $\{x_i\}^N_1$ leads to the linear system
\begin{equation}
	\mathbb{K}\,\mathbf{u}_{\rm S}(\omega)
	=  -\left[ \mathbb{I} - \mathbb{K}\,\mathbb{R}(\omega)\right]^{-1}\mathbb{K}\,\mathbb{R}(\omega)\mathbb{K}\,\mathbf{u}(\omega)\;,
\end{equation}
which can be solved for $\overline{u}_{\rm S}(x_i,\omega)$.
Here bold-face symbols denote vectors of length $N$, e.g., $\mathbf{u}=[u(x_1),\ldots,u(x_N)]$
and double struck symbols are $N\times N$ matrices. In particular, $\mathbb{I}_{ij}=\delta_{i,j}$, $\mathbb{K}_{ij}=K(x_i)\delta_{i,j}$ and $\mathbb{R}_{ij}=(2\pi)^{-2}\overline{R}_{xx}(x_i-x_j,\omega)$.
Using this result and the discretized version of the equation of motion \eqref{eq:u_eom} one obtains
an equation for the in-plane response function $\mathbb{G}_{ij}=\hat{G}(x_i,x_j,\omega)$
\begin{equation}
	\left[-\omega^2 \mathbb{I}- c^2 \mathbb{L} 
	+ \frac{1}{\rhoG} \left[ \mathbb{I} - \mathbb{K}\mathbb{R}(\omega)\right]^{-1}\mathbb{K}
	\right] \mathbb{G}(\omega)= \mathbb{I}\;,
\end{equation}
where $\mathbb{L}$ is the discrete second derivative \cite{prfl+92}. Approximating the integrations in Eq.\ \eqref{eq:ovrlp} by numerical quadratures, one finally obtains
\begin{equation}
	\FT{\chi}(\Omega) = \frac{c^2}{2}\boldgreek{\Phi}^{\rm t} \mathbb{G}(-\Omega) \boldgreek{\Phi}\;
\end{equation}
with $\boldgreek{\Phi}_i = \left.\frac{d}{dx}\ddx{\phi}^2 \right|_{x=x_i}$, which allows the computation of $\FT{\chi}$ for a given geometry. The parameters
entering the equation of motion can then be calculated using Eqs.\ \eqref{eq:env_coeff}.
Following Ref.\ \citenum{licr08}, we set
$\tilde{\gamma} = \gamma/(m \Omega_0)$,
$\tilde{\eta} = \eta \Omega_0/\alpha$,
$\tilde{g} = g \sqrt{\frac{\alpha}{m^3}}/\Omega_0^3$,
$\tilde{\Omega} = \Omega/\Omega_0$, and $\tilde{q} = q\,\sqrt{\alpha/m \Omega_0^2}$.
In the limit of weak LD, $\tilde{\gamma}\ll 1$, the response of the resonator 
is determined solely by the dimensionless parameters $\tilde{\eta}$, $\tilde{g}$ and $\tilde{\Omega}$, describing the nonlinear damping, the driving strength and the driving frequency.

\section{Results}\label{sec:results}
To quantify the influence of LD and NLD, we consider the setup shown in
Fig.\ \ref{fig:setup} with a back-gate voltage 
$V_{\rm bg} = V_{\rm dc} + V_{\rm ac}\cos(\Omega t)$.
The fundamental-mode shape is taken to be $\phi(x) = \sqrt{2} \cos(\pi x/\ell)$, 
which gives $\alpha_0 = 3 T_1 \pi^4 \bee/(4 \ell^3)$. 
Within a parallel plate model for electrostatic actuation, the
force acting on the graphene sheet is given by
\begin{align}
f(x) ={}& \frac{\partial}{\partial w}\frac{1}{2} C(w) V_{\rm bg}^2 \notag\\
{}&\approx -\frac{\epsilon_0 }{2(d + q(t)\phi(x))^{2}} \left( V_{\rm dc}^2 + 2 V_{\rm dc}V_{\rm ac}\cos(\Omega t)\right)\;,\label{eq:force}
\end{align}
where $C(w)=\epsilon_0/(d+w)$ is the capacitance of a parallel plate capacitor with plates being separated by 
the distance $d+w$ and $\epsilon_0$ is the vacuum permittivity. The distance is determined by the depth $d$ of the trench and the flexural displacement
$w$ of the resonator. In the second line we further assumed $V_{\rm dc}\gg V_{\rm ac}$, which
is typically found in experiments. The force can be separated into
a static and a time-dependent part, $f=f_{\rm dc} + f_{\rm ac}\cos(\Omega t)$ with $f_{\rm dc}\propto V_{\rm dc}^2$
and $f_{\rm ac}\propto V_{\rm dc}V_{\rm ac}$, respectively. Since the displacement, which is
on the order of a few nanometers, is much smaller than the trench depth, the force can be expanded in powers of
$w$.
Accordingly, the driving strength in Eq.\ \eqref{eq:env_eom} becomes
$g = 2 \sqrt{2} \ell \bee \epsilon_0 V_{\rm dc} V_{\rm ac}/(\pi d^2)$. Moreover,
the static displacement can be found by solving Eqs.\ \eqref{eq:w_eom} and 
\eqref{eq:u_eom} in the static limit (see Appendix \ref{sec:app_static}). This yields $q_0 \approx \sqrt{2} \ell^2 \epsilon_0 V_{\rm dc}^2/(\pi^3 d^2 T_0)$. Note the dependence on the tensile stress $T_0$; $q_0$ becomes
smaller for increasing tensile stress.


	\begin{table}[tb!]
	\begin{tabular*}{\columnwidth}{@{\extracolsep{\fill}}lcc}
		\multicolumn{3}{c}{graphene and substrate parameters} \\
		\hline
		\hline
		graphene mass density  & $\rhoG$ & $7.6\times 10^{-7} {\rm \;kg\;m^{-2}}$ \\
		$\lambda_{\rm G}+2\mu_{\rm G}$				  & $T_1$ & $340 {\rm \;N\;m^{-1}}$ \\
		SiO${}_2$ mass density  & $\rhoS$ & $2.2\times 10^{3} {\rm \;kg\;m^{-3}}$ \\
		SiO${}_2$ sound velocities & $c_{\rm L}/c$ & $0.28$\\
							  & $c_{\rm T}/c$ & $0.18$\\
		coupling strength & $K_0$ & $1.82\cdot 10^{20} \,{\rm N\;m^{-3}}$\\
		\hline		
		\multicolumn{3}{c}{resonator parameters} \\
		\hline
		\hline
		total	length			& $L$ & $2 {\rm \;\mu m}$\\
		length			& $\ell$ & $1 {\rm \;\mu m}$\\
		width			& $\bee$ & $1 {\rm \;\mu m}$\\
		distance to gate	& $d$	&	$330 {\rm \;nm}$\\
		tensile stress & $T_0$ & $0.34 {\rm \;N\;m^{-1}}$	\\
		\hline
	\end{tabular*}
	\caption{Graphene and resonator parameters used for the calculations in Figs.\ \ref{fig:lin-nonlin} and \ref{fig:VdcVacQfactor}. Graphene and substrate parameters are taken from Refs.\ \citenum{lewe+08} and \citenum{peue10}.}
	\label{tab:params}
	\end{table}
In the following, we consider a graphene resonator with dimensions and 
parameters as given in Tab.\ \ref{tab:params}. We checked that the results do not change, for
larger values of the total length $L$.
Using Eqs.\ \eqref{eq:env_coeff} and \eqref{eq:teta} we obtain $\alpha/\alpha_0 \approx 0.64$ and $\tilde{\eta}\approx7\cdot10^{-4}$. 
The latter implies bi-stable behavior of the resonator. 
In general, these values depend sensitively on the geometry of the graphene sheet and
on the substrate. Our results provide a ``best case'' estimate, since the substrate is
treated as a semi-infinite medium and the trench is modeled by the position dependent
coupling $K(x)$. Lifting these restrictions will lead to a stronger response of the 
substrate, and more dissipation. 

\subsection{Resonance frequency}\label{sec:res_freq}
As described in Sec.\ \ref{sec:eom} the resonance frequency $\Omega_0(V_{\rm dc})$ depends on the initial
stress and the bias voltage. The dependence of $\Omega_0$ on bias voltage, the so called tuning curve, is a characteristic feature of NEMS devices. It is a result of the competition between softening (decreasing $\Omega_0$) due to the electrostatic force [Eq.\ \eqref{eq:force}], and stiffening (increasing $\Omega_0$) due to the Duffing nonlinearity of the graphene sheet. 

To obtain the tuning curve, we separate static and dynamic contributions to the displacement fields, 
\begin{subequations}
\begin{align}
w(x,t)&=w_0(x) + \delta w(x,t)\;, \\
u(x,t)&=u_0(x) + \delta u(x,t)\;
\end{align}
\end{subequations}
and insert these expressions into the equations of motion given by Eqs.\ \eqref{eq:eom}. The static
solutions, $w_0$ and $u_0$, are calculated in Appendix \ref{sec:app_static}. Further,
we expand the static force $f_{\rm dc}(x)$ up to first order in $\delta w$,
\begin{equation}
f_{\rm dc}\approx-\frac{\epsilon_0 V_{dc}^2}{2(d+w_0)^2}+\frac{\epsilon_0 V_{dc}^2}{(d+w_0)^3}\delta w\;.
\end{equation}
The resonance frequency is then obtained by collecting terms, which are linear in the
vibration amplitude $\delta w$. There are three such terms, which contribute to the resonance frequency,
\begin{subequations}\label{eq:res_tuning}
\begin{equation}
	\Omega_0^2(V_{\rm dc}) = \Omega_0^2(0) + \Delta\Omega_{\rm mech.}^2 - \Delta\Omega_{\rm el.}^2\;
\end{equation}
with
\begin{align}
	\Omega_0^2(0) &= \frac{T_0}{\rhoG}\frac{\pi^2}{\ell^2}\;, \\
	\Delta\Omega_{\rm mech.}^2(V_{\rm dc}) &= 2\frac{T_1 \pi^4}{\rhoG \ell^4} q_0^2=\frac{8}{3 m}\alpha_0 q_0^2\;, \\
	\Delta\Omega_{\rm el.}^2(V_{\rm dc}) &= \frac{\epsilon_0 V_{\rm dc}^2}{d^3 \rhoG}\;.
\end{align}
\end{subequations}
The three contributions are due to initial strain, mechanical stiffening and electrostatic softening,
respectively. Since the static deflection $q_0$ depends on the bias voltage $V_{\rm dc}$, the last two terms
yield the voltage dependent tuning behavior.

\begin{figure}[t!]
	\includegraphics[width=.95\columnwidth]{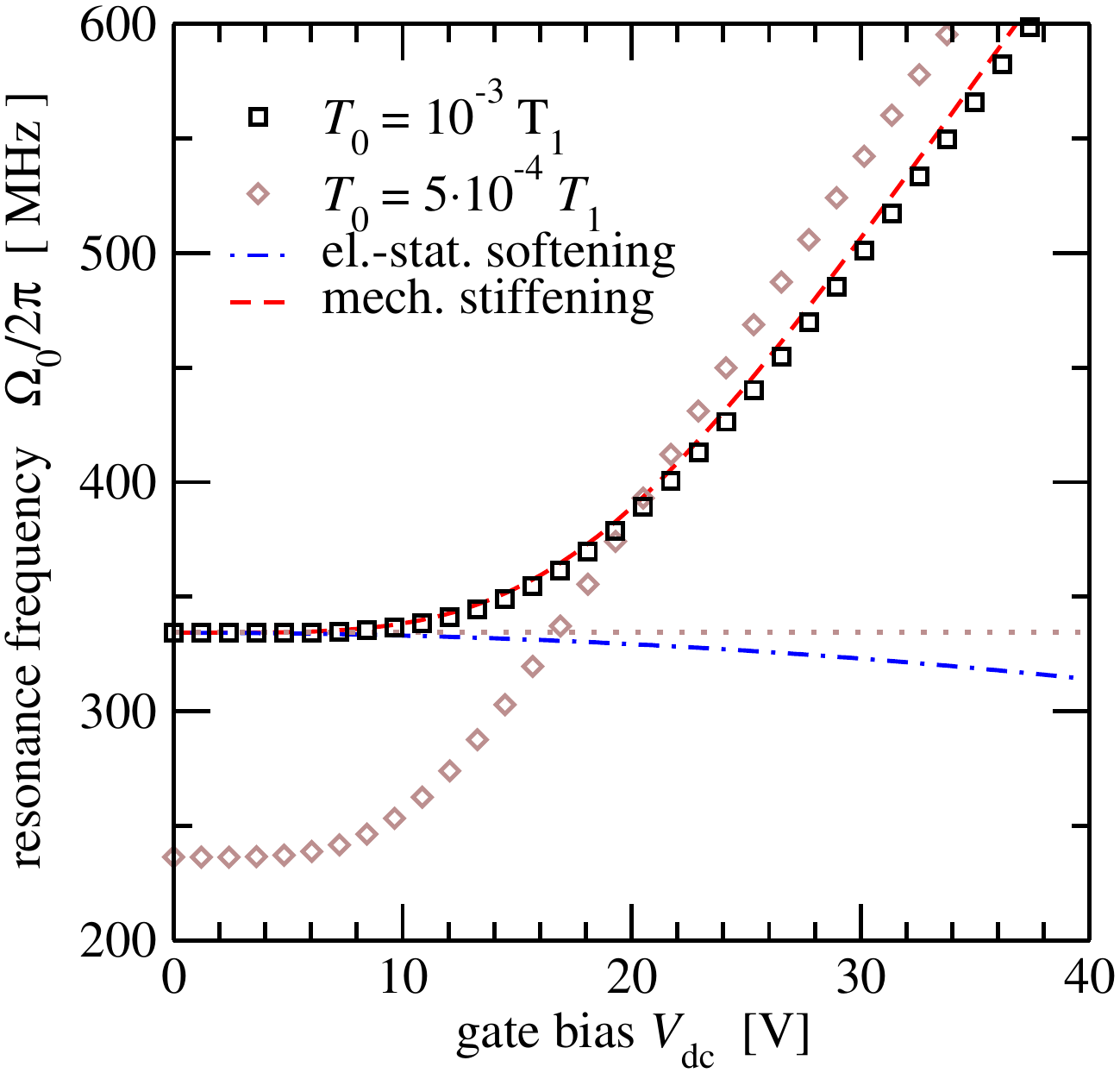}
	\caption{(Color online) Resonance frequency $\Omega_0$ {\em vs.}\ bias voltage.
	Symbols denote results of numerical calculation. 
	The dashed (red) and dashed-dotted (blue) lines show the contributions of
	mechanical stiffening and electrostatic softening for $T_0=10^{-3} T_1$, respectively. 	
	Parameters are given in Tab.\ \ref{tab:params}.}
	\label{fig:res_tuning}
\end{figure}
Figure \ref{fig:res_tuning} shows the tuning curve for the parameters given in in Tab.\ \ref{tab:params}.
For voltages, $V_{\rm dc} > 10\;\text{V}$, the resonance frequency (squared) is mainly determined by the mechanical stiffening, which scales with $V_{\rm dc}^4$ while the softening term scales with $V_{\rm dc}^2$ according Eqs.\ \eqref{eq:res_tuning}.

Depending on the specific geometry and the initial stress, the resonance frequency of
the resonator may be substantially tuned using the bias voltage. Since the linear and nonlinear 
damping coefficients given by Eqs.\ \eqref{eq:env_coeff} depend on frequency, the magnitude of
LD and NLD will, in principle, also be influenced by the tuning curve. In order to disentangle the influence
of $\Omega_0(V_{\rm dc})$ and the coupling to the in-plane phonons, we will
only consider a constant resonance frequency $\Omega_0=\Omega_0(0) = \sqrt{T_0/\rhoG}(\pi/\ell)$
in the following discussions (see Appendix \ref{sec:inf_tuning} for the influence of
the tuning on the quality factor).

\subsection{Damping ratio}\label{sec:damping_ratio}
The relative importance of LD and NLD, which is quantified 
by $\tilde{\delta}$ defined in Eq.\ \eqref{eq:lin-nonlin},
is determined by the ratios $\Im\FT{\chi}(2\Omega)/(8\Im\FT{\chi}(\Omega))$ and $|q^{\rm max}_1|/q_0$. The former weakly depends  on the geometric details.
%
	\begin{figure}[t!]
		\includegraphics[width=.99\columnwidth]{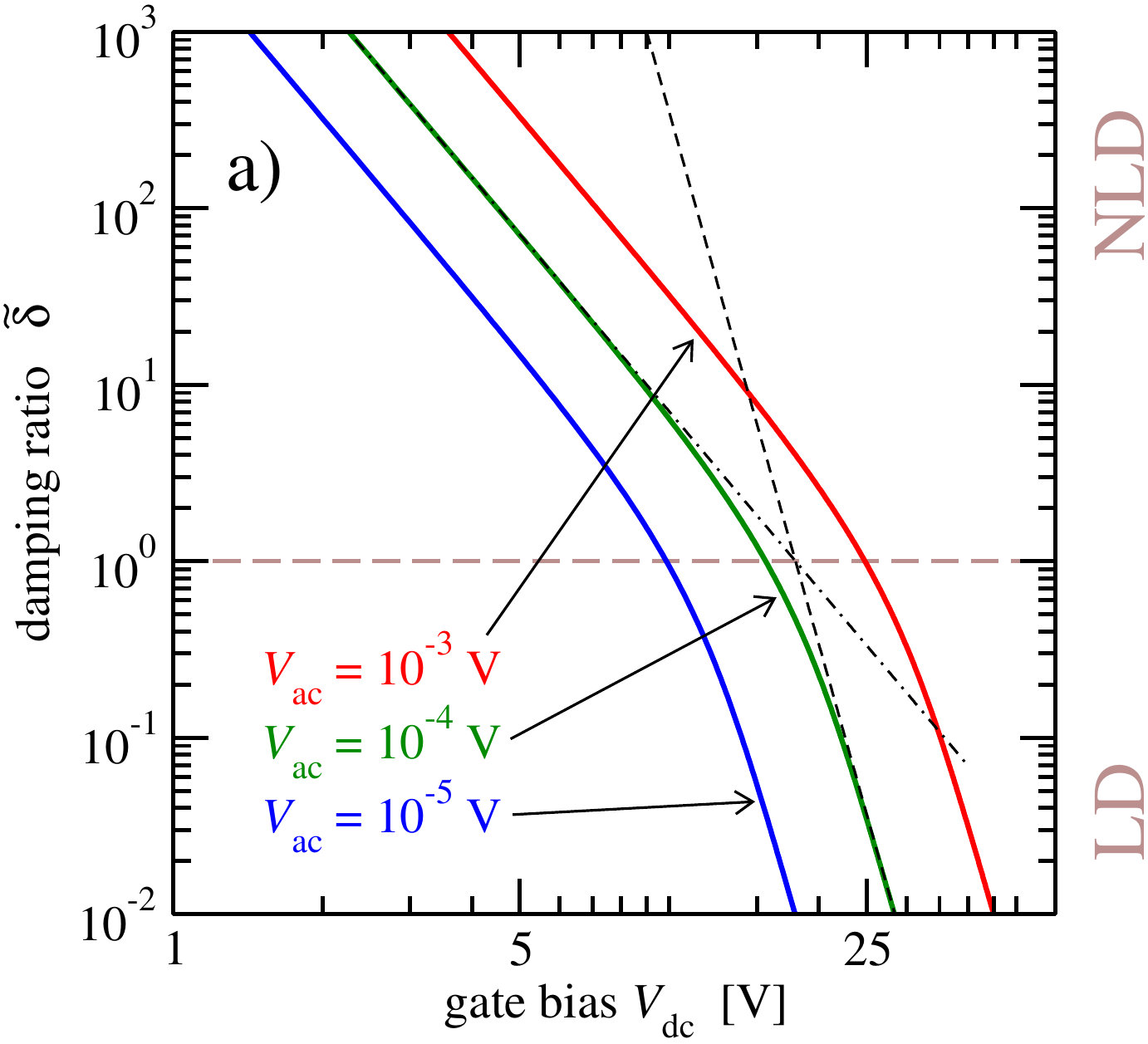}
		\includegraphics[width=.99\columnwidth]{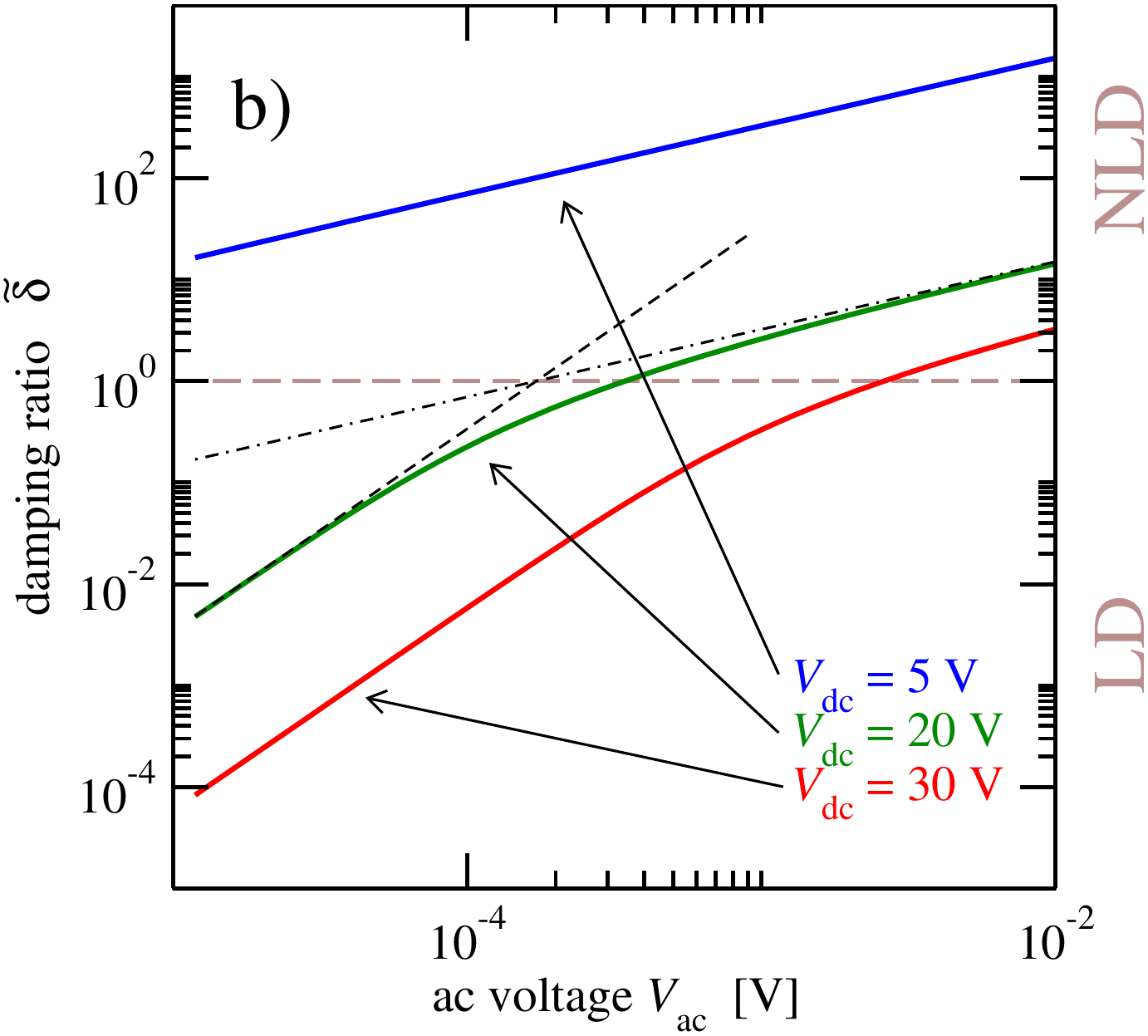}		
		\caption{(Color online) Ratio $\tilde{\delta}$ of nonlinear (NLD) and 
		linear (LD) damping 
		terms according to 
		Eq.\ \eqref{eq:lin-nonlin}; a) bias voltage and b) ac voltage dependence.
		The thin dashed and dashed-dotted lines show the asymptotic behavior
		for strong LD and NLD. A crossover between the two regimes is achieved by 
		changing the bias voltage.
		Parameters are given in Tab.\ \ref{tab:params}.}
		\label{fig:lin-nonlin}
	\end{figure}
	For small $\Omega$ one can expand $\Im\FT{\chi}(\Omega)$ in odd powers of $\Omega$.
	As $\Im\FT{\chi}$ is proportional to the density of states of the substrate phonons,
	$D(\Omega)\propto \Omega$, we expect on symmetry-grounds for a quasi-1D geometry, that
	$\Im\FT{\chi}(\Omega)\propto \Omega^3$.
	Consistent with this expectation, we obtain numerically $\Im\FT{\chi}(2\Omega)/(8\Im\FT{\chi}(\Omega))\approx 0.93$.
	
	The maximum amplitude $q_1^{\rm max}$ can easily be found from Eq.\ \eqref{eq:env_eom}
	in the steady-state limit, which yields an implicit equation for the magnitude $|q_1|$ of the steady-state 
	amplitude \cite{licr08}. Sweeping the driving frequency, the maximum amplitude 
	is attained when $d|q_1|/d\Omega = 0$, which results in the cubic equation
	\begin{equation}
		4 \tilde{g} = |\tilde{q}_1^{\rm max}| (4 \tilde{\gamma} + \tilde{\eta}|\tilde{q}_1^{\rm max}|^2)\;.
	\end{equation}
	Here, $\tilde{\gamma}$ and $\tilde{g}$ depend on the bias voltage $V_{\rm dc}$
	via $q_0$ and $f_{\rm ac}$, respectively. However, note that only $\tilde{g}$ depends
	on the ac voltage.
	Due to the different dependencies of $q_0$ and $|q_1^{\rm max}|$ on the bias
	voltage, one can achieve a crossover from NLD to LD
	dominated behavior by increasing the bias voltage.
	This is shown in Fig.\ \ref{fig:lin-nonlin}a. In the limit of small $V_{\rm dc}$,
	$|\tilde{q}_1^{\rm max}| \approx \left( 4 \tilde{g}/\tilde{\eta} \right)^{1/3}
	\propto V_{\rm ac}^{1/3}V_{\rm dc}^{1/3}$ and $\tilde{\delta}> 1$, {\it i.e.}, NLD dominates.
   For large $V_{\rm dc}$, 
	$|\tilde{q}_1^{\rm max}| \approx \tilde{g}/\tilde{\gamma}
	\propto V_{\rm ac}V_{\rm dc}^{-3}$ and $\tilde{\delta}$ goes to zero with increasing
	$V_{\rm dc}$.
	Since the static displacement is determined only by the
	geometry and the bias voltage, and the maximal amplitude additionally depends
	on the ac voltage, the crossover can also be realized by tuning $V_{\rm ac}$, which
	is shown in Fig.\ \ref{fig:lin-nonlin}b.
	Equating the expressions for $|\tilde{q}_1^{\rm max}|$ in the two limits gives 
	an estimate for the crossover for both voltages.
	Additionally, due to the dependencies of $q_0\propto T_0^{-1}$ and $\Omega_0\propto\sqrt{T_0}$
	on the initial tension $T_0$
	one finds that the damping ratio $\tilde{\delta}$ increases with increasing tension
	in both regimes ($\tilde{\delta} \propto T_0^3$ and $\tilde{\delta}\propto T_0$ 
	in the LD and NLD regime, respectively). Thus, the non-linear damping is enhanced for larger $T_0$.

	\subsection{Quality factor}\label{sec:quality}
	To quantify the energy loss we consider
	the quality factor $Q=\Omega_0 \mean{E_\perp}/\mean{\dot{E}_\perp}$, 
	which measures the 
	time-averaged dissipated energy $\mean{\dot{E}_\perp}$ normalized to
	the average energy $\mean{E_\perp}$ in the flexural modes. 
	The nonlinearities render $Q$ amplitude dependent. To get a worst case estimate, we use
	the maximal amplitude. In the slow envelope approximation we find
	\begin{equation}\label{eq:Qfactor}
		\frac{1}{Q}
		\approx \frac{\Omega_0 \left(\gamma + \frac{1}{4} \eta |q_1^{\rm max}|^2\right)}{m \Omega_0^2 + \frac{1}{2}\frac{3}{8} \alpha |q_1^{\rm max}|^2} \;.
	\end{equation}
	The nature of the damping influences $Q$. In the LD dominated regime, 
	$\tilde{\delta}\ll 1$, $Q$ is independent of the vibrational amplitude, 
	$Q_{\rm LD}\approx m \Omega_0/\gamma$. In contrast, for
	$\tilde{\delta}>1$ one gets 
	$Q_{\rm NLD}\approx 4 m \Omega_0/(\eta |q_1^{\rm max}|^2)$
	for $\tilde{\eta}>1$. Thus, $Q$ increases with decreasing driving strength.
	This agrees with the conclusions of Ref.\ \citenum{eimo+11}.

\begin{figure}[t!]
		\centering
		\includegraphics[width=.9\columnwidth]{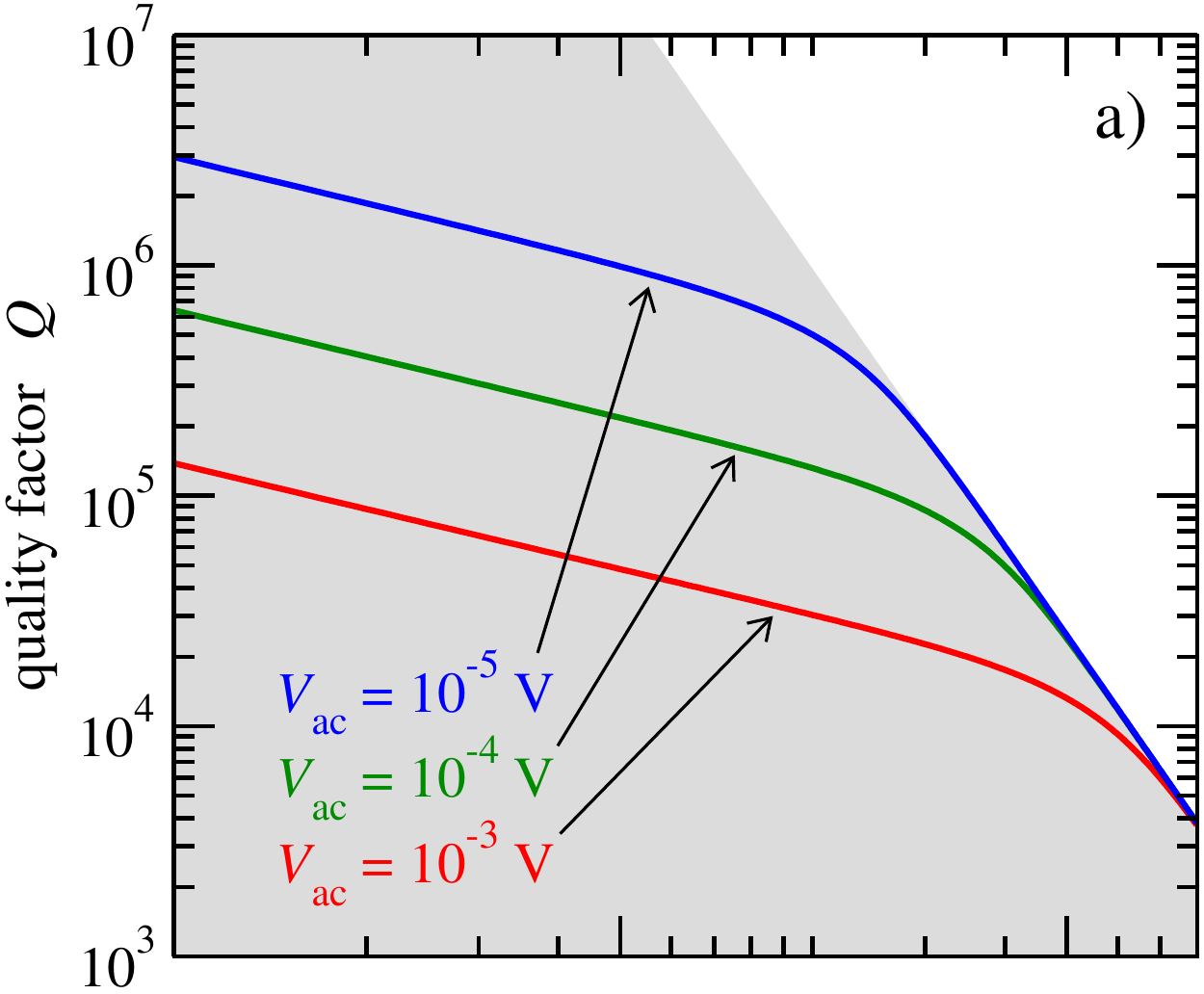}\hfill
		\includegraphics[width=.9\columnwidth]{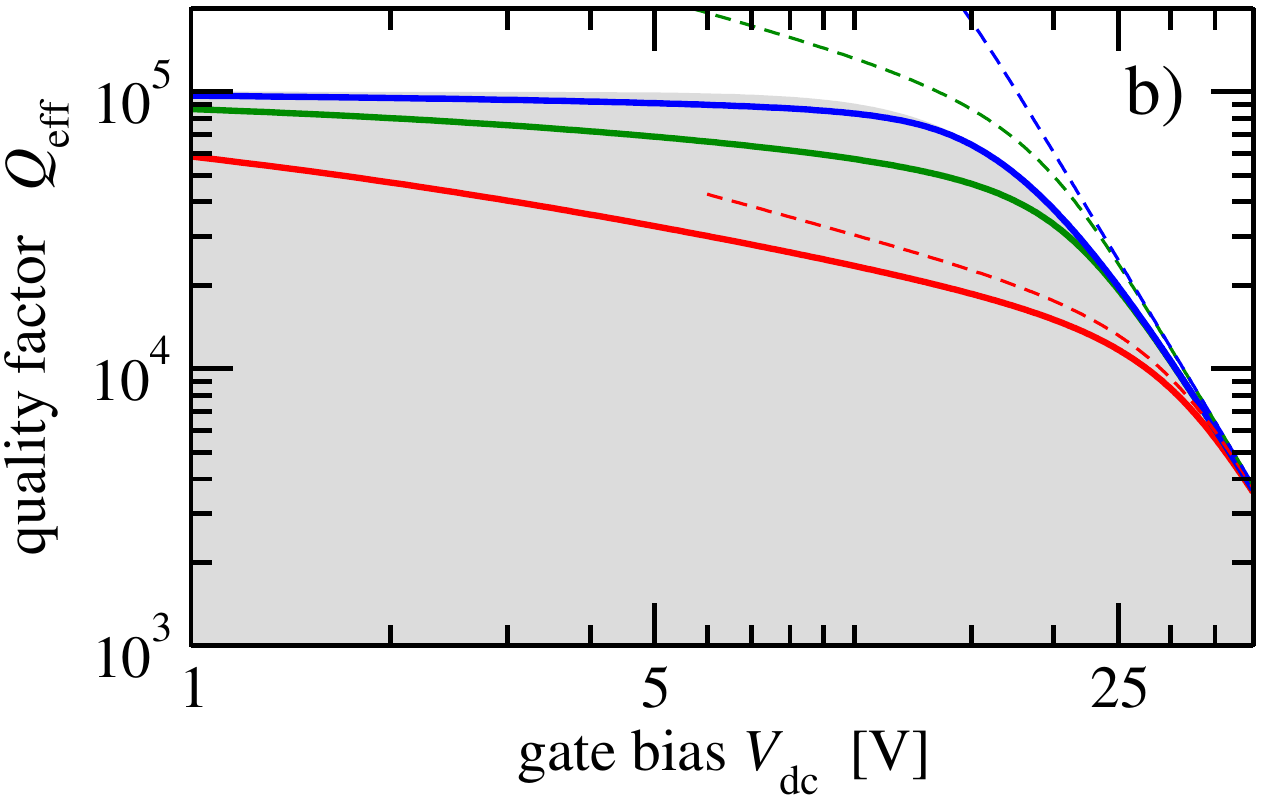}
		\caption{(Color online) Quality factor $Q$ {\em vs.}\ bias
		voltage. a) $Q$ calculated from Eq.\ \eqref{eq:Qfactor} and b) 
		with additional voltage independent damping, $Q_{\rm eff}^{-1}=Q^{-1} + Q_0^{-1}$
		with $Q_0=10^5$. The gray area indicates the region of
		attainable $Q$-factors. The dashed lines correspond to the behavior in a).
		Parameters are given in Tab.\ \ref{tab:params}.}
		\label{fig:VdcVacQfactor}
	\end{figure}

	Figure \ref{fig:VdcVacQfactor}a shows the quality factor 
	as a function of bias voltage for constant
	$V_{\rm ac}$.
	As expected,
	$Q$ decreases with increasing bias and excitation voltages and its behavior with regard to applied voltage
   changes qualitatively at the crossover between LD and NLD regimes. The asymptotic LD
   behavior limits the maximally attainable $Q$-factor, which is indicated by the
   gray area. 
   We also compare to the case where the LD is additionally caused by a mechanism 
   that does not 
	depend on the bias voltage leading to $Q_0$. In this case the effective
	$Q$-factor, $Q_{\rm eff}^{-1}=Q^{-1} + Q_0^{-1}$, has a cutoff for small $V_{\rm dc}$
	as shown in Fig.\ \ref{fig:VdcVacQfactor}b, which further limits the region of attainable
	$Q$-factors. The qualitative difference between LD and NLD is still present
	and should be experimentally observable. Most importantly, by decreasing $V_{\rm ac}$
	the maximally attainable $Q$-factor, which is determined by other damping mechanisms
	can be approached.

\section{Conclusions}			
	In conclusion, we have studied coupling between flexural vibrations and in-plane 
	displacements as a physical mechanism for
	damping of flexural modes in graphene resonators. A characteristic
	consequence, which influences the behavior of
	the dependence of the quality factor on bias and excitation voltages,
	is the competition between static deflection and 
	vibrational amplitude.
	We note that the same type of behavior would naturally occur for any dissipative
	process which couples linearly to the strain; for example, Ohmic dissipation induced
	by synthetic gauge fields \cite{vogu+09}.
	The cross-over should allow for an experimental verification of this class of damping 
	mechanisms.

\begin{acknowledgements}
	We thank J.\ Atalaya for helpful discussions.
	The research leading to these results has received funding [DM,AI] from the
	EU $7^{\rm th}$ framework program (FP7/2007-2013) RODIN (grant agreement no: 246026)
	and the Swedish Research Council [JK].
\end{acknowledgements}

\appendix

\section{Response of an elastic half-space}\label{sec:app_response}
The displacement response at the surface of an elastic half-space to
a stress acting on the surface is given in terms of a response function by Eq.\ \eqref{eq:sub_resp}.
If the stress is directed parallel to the $x$-axis, the spatial Fourier transform
of Eq.\ \eqref{eq:sub_resp} reads
\begin{equation}
	u_{\rm S}(\vec{k},z=0,\omega) = - R_{xx}(\vec{k},\omega) \sigma_{x z}(\vec{k},\omega)
		\;,
\end{equation}
where $\vec{k}=(k_x,k_y)$ is the surface wave vector.
The response function $R_{xx}(\vec{k},\omega)$ for finite frequencies 
is explicitly given by\cite{pe01,mami76}
\begin{subequations}
\begin{equation}
	R_{xx}(\vec{k}, \omega) 
	= -\frac{\imath}{\rho_{\rm S} \cT^2}
		\left(
			\frac{p_{\rm T}(\omega, k)}{S(\omega, k)} \frac{\omega^2}{\cT^2}\frac{k_x^2}{k^2}
			+\frac{1}{p_{\rm T}(\omega, k)} \frac{k_y^2}{k^2}
		\right)
\end{equation}
with
\begin{align}
	p_{\rm L,T}(\omega, k) &= \sqrt{\left(\frac{\omega}{c_{\rm L,T}}\right)^2 +\imath \varepsilon - k^2} \;,\\
	S(\omega, k) &=	\left[ \left(\frac{\omega}{c_{\rm L,T}}\right)^2 - 2 k^2 \right]^2
						+ 4 k^2 p_{\rm L}(\omega, k) p_{\rm T}(\omega, k)\;,
\end{align}
\end{subequations}
where $\cL$ and $\cT$ are the longitudinal and transversal speeds of sound, respectively, and the infinitesimal
$\varepsilon>0$ ensures causality.
Notice that $p_{\rm L,T}$ and $S(\omega, k)$ depend only on the modulus $k$ of the wave vector $\vec{k}$.
The response function in real space is then
\begin{subequations}\label{eq:sub_resp_func_I}
\begin{align}
	R_{xx}(\vec{x}, \omega) &= \int d^2k R_{xx}(\vec{k}, \omega) e^{\imath \vec{k} \cdot \vec{x}} \notag\\
	&= -\frac{2\pi \imath}{\rho_{\rm S} \cT^2} 
	\left(
	\frac{\partial}{\partial x}
	I_{x}(x, y)
	+ 
	\frac{\partial}{\partial y}
	I_{y}(x, y)
	\right) \;. 
\end{align}
Here, we defined
\begin{align}
	I_{x}(x, y) &= \frac{x}{\sqrt{x^2+y^2}} \left(\frac{\omega}{\cT}\right)^2  \int dk\;
			\frac{p_{\rm T}(\omega, k)}{S(\omega, k)} J_1( k \sqrt{x^2+y^2}) \;,\\
	I_{y}(x, y) &= \frac{y}{\sqrt{x^2+y^2}} \int dk\; \frac{1}{p_{\rm T}(\omega, k)}J_1( k \sqrt{x^2+y^2})\;,
\end{align}
\end{subequations}
where $J_1$ is a first order Bessel function of the first kind.
Note, that
\begin{equation}
	I_{x}(x, -y) = I_{x}(x, y)\,,\quad
	I_{y}(x, -y) = -I_{y}(x, y)\;.
\end{equation}
The expressions given in Eqs.\ \eqref{eq:sub_resp_func_I} are a very convenient starting
point for the numerical evaluation of the response function used in Sec.\ \ref{sec:num_method}.

The zero-frequency response can be directly calculated in real space\cite{lali86}. One finds
\begin{equation}\label{eq:stat_resp_func}
	R_{xx}(\vec{x}, \omega=0)
	= \frac{1}{4\pi \rhoS \cT^2}\frac{2(\cT^2-\cL^2) x^2 - \cL^2 y^2}{(\cL^2-\cT^2)(x^2+y^2)^{3/2}}\;.
\end{equation}

\section{Static displacement}\label{sec:app_static}
In the static limit, the equations for the in-plane and out-of-plane displacements \eqref{eq:eom} 
within the suspended region become
\begin{subequations}
\label{eq:static}
\begin{align}
\label{eq:staticu}
T_1 u_{,xx} +  \frac{T_1}{2}\partial_x \left(\ddx{w}^2\right)=&0\;,\\
-\frac{T_1}{2}\partial_x \left[\left(2 \ddx{u}+\ddx{w}^2\right)\ddx{w}\right]=&f_{dc}\label{eq:staticw}\;,
\end{align}
\end{subequations}
with vanishing boundary conditions at $x=\pm\ell/2$ for the out-of-plane displacement. 
To find the proper boundary conditions for the in-plane displacement, we need to consider the coupling to the substrate in the non suspended region. Here, the equation for the in-plane displacement \eqref{eq:u_eom} 
is given by
\begin{equation}
\label{eq:inplaneclamped}
T_1 u_{,xx}-K(x)(u(x)-\overline{u}_S/b)=0\;.
\end{equation}
Following the same line of reasoning as in the main text, the static substrate response can be written as
\begin{align}
\overline{u}_{\rm S}(x)  ={}&  -\int\limits^{L/2}_{-L/2} \frac{d x'}{(2\pi)^2} 
		\overline{R}_{xx}(x-x')\,\Theta(|x'|-\ell/2) h(x')
	   \;. \label{eq:sub_resp_barapp}
\end{align}
with $h(x)=K_0(u-\overline{u}_S/\bee)$ and $\overline{R}_{xx}(x-x')$ being the static response function 
for an elastic half space given by Eq.\ \eqref{eq:stat_resp_func}
integrated over $y$. To treat the problem analytically, we convert Eqs.\ \eqref{eq:inplaneclamped} and \eqref{eq:sub_resp_barapp} into a local equation for the in-plane displacement. In the limit of very strong coupling to the substrate, the spatial variation of $h(x)$ is small, in which case
\begin{align}
-&\int\limits^{L/2}_{-L/2} \frac{d x'}{(2\pi)^2} 
		\overline{R}_{xx}(x-x')\, \Theta(|x'|-\ell/2)h(x')\approx \nonumber \\
&-h(x)\int\limits^{L/2}_{-L/2}\frac{d x'}{(2\pi)^2} 
		\overline{R}_{xx}(x-x') \Theta(|x'|-\ell/2)\;.
\end{align}
This makes it possible to solve for $h(x)$ in terms of the in-plane displacement $u(x)$. One finds
\begin{equation}
h(x)=\frac{K_0}{1-\overline{R}_0(x)K_0}u(x)\;,
\end{equation}
where $\overline{R}_0(x)\equiv(2\pi)^{-2}\int\limits^{L/2}_{-L/2} d x'\; \overline{R}_{xx}(x-x') \Theta(|x'|-\ell/2)$.
This expression is valid outside the suspended region and is approximately given by $h(x)\approx -1/\overline{R}_0(x)$, which assumes $K_0 \overline{R}_0(x)\gg 1$. Consequently, the equation for the in-plane displacement, Eq.\ \eqref{eq:inplaneclamped}, is modified to become
\begin{equation}
T_1 u_{,xx}+\overline{R}_0(x)^{-1} u = 0
\end{equation}
for $|x|>\ell/2$. Thus, the effect of the substrate is reduced to that of a spring with a spatially varying spring constant. The displacement $u$ is expected to decay exponentially to zero in the clamped region with 
a decay length $\lambda\equiv \sqrt{\overline{R}_0(x) T_1}$. For the substrate parameters given in Table \ref{tab:params}, this amounts to $\lambda\approx 100$ nm. As a consequence, within a distance of $100$ nm from the edge of the suspended region the in-plane displacement $u(x)$ is essentially zero. To a good approximation, we therefore assume vanishing boundary conditions for in-plane displacement at $|x|=\ell/2$. 

Setting $u(x)=(T_0/T_1)x+\Delta u(x)$, where the first terms accounts for initial strain in the graphene, the boundary conditions are $w(x=\pm \ell/2)=0$ and $\Delta u(x=\pm \ell/2)=0$. Using the Ansatz $w(x)=q_0 \phi(x)$ with 
$\phi(x)=\sqrt{2}\cos\pi x/\ell$, the in-plane equation \eqref{eq:staticu} reads
\begin{equation}
\Delta u_{,xx}=-\frac{q_0^2}{2}\partial_x \left(\ddx{\phi}^2\right)\;.
\end{equation}
Consequently, the in-plane displacement will be given by
\begin{equation}
\Delta u(x)= -q_0^2\frac{\pi^2}{\ell^2}\int\limits^{x}_{0} dx' \sin^2\pi x'/\ell + \frac{\pi^2}{2\ell^2}q_0^2 x\;.
\end{equation}
Inserting this expression into Eq.\ \eqref{eq:staticw} and we obtain
\begin{equation}
q_0\left( \frac{\pi^2}{\ell^2} T_0+  \frac{\pi^4}{2 \ell^4} T_1 q_0^2\right) =\frac{2\sqrt{2}}{\pi}f_{dc}\;.
\end{equation}
This is a purely algebraic equation for the static deflection. In the limit $q_0\ll \frac{\ell}{\pi}\sqrt{\frac{T_0}{T_1}}\approx 10$ nm for $\ell=1\mu$m and $T_0/T_1=10^{-3}$,Êthe cubic term can be 
neglected and $q_0\propto f_{\rm dc}$.
\begin{figure}[t!]
 \centering
 \includegraphics[width=.9\columnwidth]{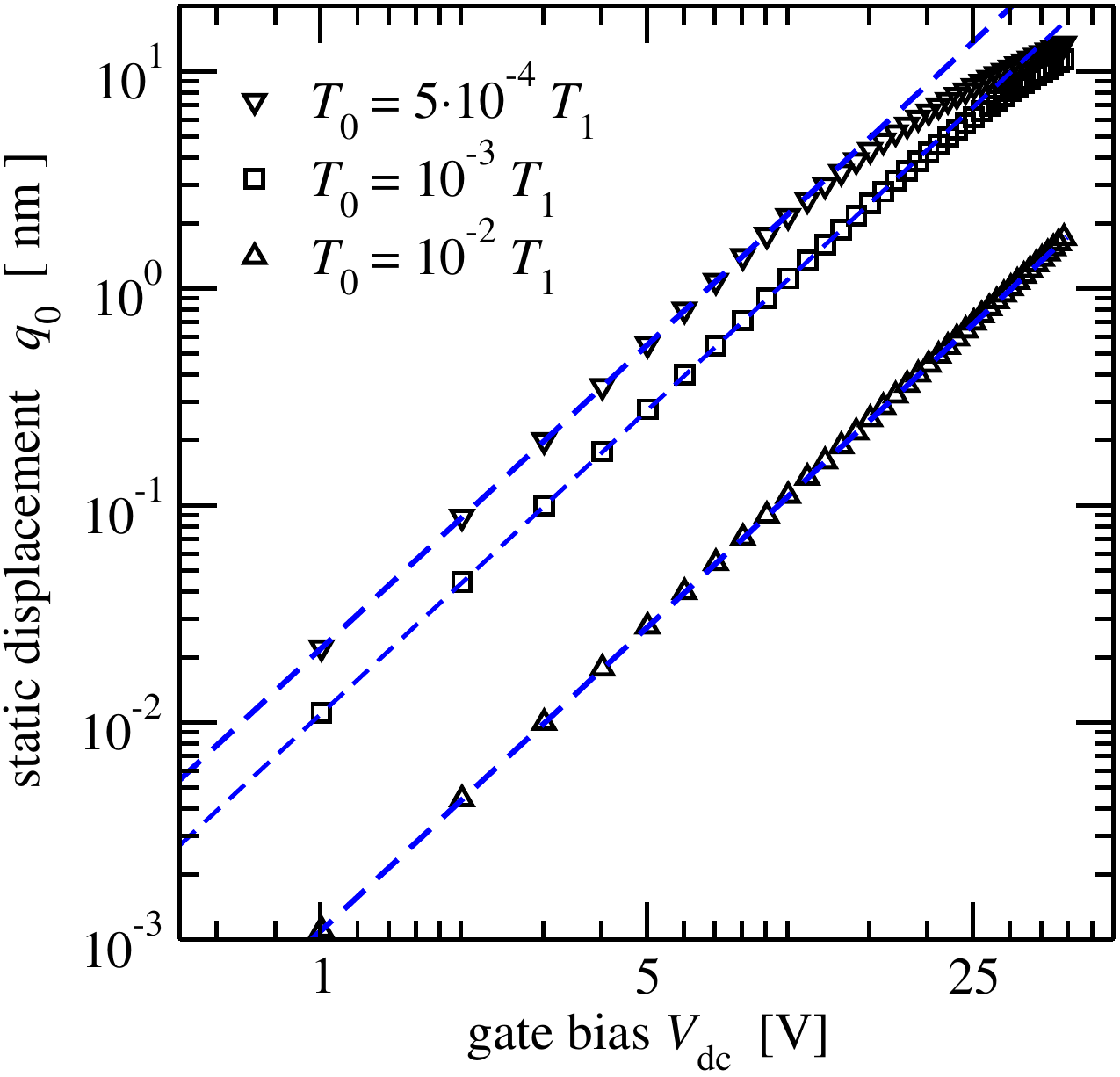}
   \caption{Static deflection $q_0$ {\em vs.} bias voltage for three values of initial tension. The linear approximation [Eq.\ \eqref{eq:q0lin}] is shown as dashed lines in the figure, while the squares and triangles correspond to the full numerical solution of the static problem.}\label{fig:q0Comp}
\end{figure}

To compute $q_0$, we need to consider the electrostatic interaction with the back gate. 
The static force acting on the graphene is given by Eq.\ \eqref{eq:force}.
Considering the limit $q_0\ll d$, we obtain for the static displacement
\begin{equation}
\label{eq:q0lin}
q_0=-\sqrt{2}\frac{\ell^2\epsilon_0 V_{dc}^2}{\pi^3 T_0d^2}\;,
\end{equation}
which is the expression given in Sec.\ \ref{sec:results}.
In Fig.\ \ref{fig:q0Comp} the linear approximation (dashed line), given by Eq.\ \eqref{eq:q0lin}, is compared to the full numerical solution of Eq.\ \eqref{eq:static} (squares and triangles), which takes the substrate into account. The linear approximation remains valid in the displayed interval for the two larger values of initial strain $T_0/T_1$, while a more significant deviation is apparent for the lowest value of the strain.

\section{Influence of tuning and initial tension on the quality factor}\label{sec:inf_tuning}
\begin{figure}[t!]
 \centering
 \includegraphics[width=.9\columnwidth]{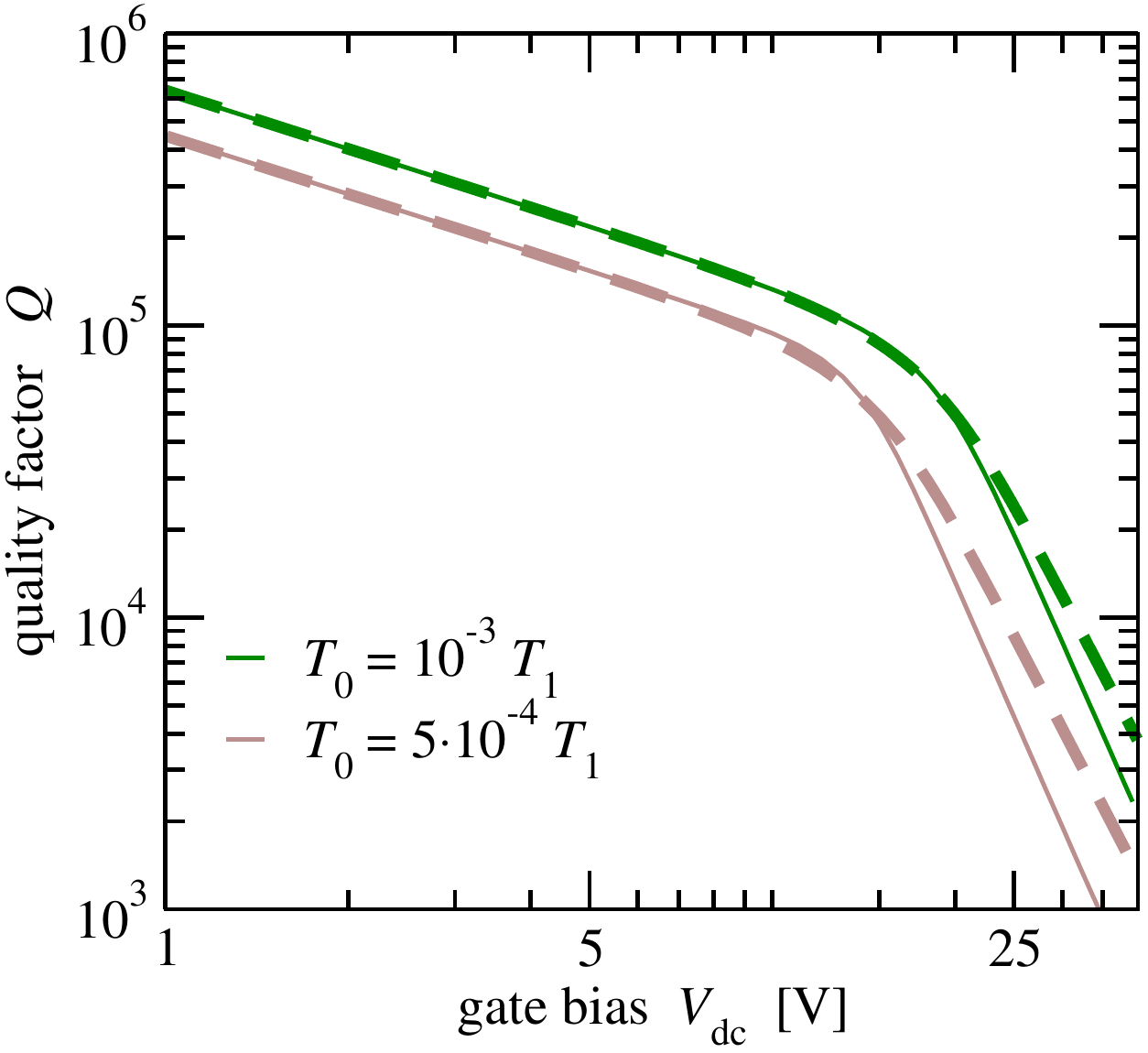}
   \caption{(Color online) Quality factor $Q$ {\em vs.}\ bias
		voltage calculated from Eq.\ \eqref{eq:Qfactor} for
		$V_{\rm ac}=10^{-4}\;\text{V}$.
		The full and dashed lines
		show the result for a voltage dependent $\Omega_0(V_{\rm dc})$ and constant $\Omega_0=\Omega_0(0)$,
		respectively.
		Parameters are given in Tab.\ \ref{tab:params}.}\label{fig:Qtuning}
\end{figure}
In Sec.\ \ref{sec:res_freq} we discussed the voltage dependence of the resonance frequency
(tuning curve) and showed that the frequency can be substantially tuned by changing
the bias voltage $V_{\rm dc}$. Since the linear and nonlinear damping constants
given by Eqs.\ \eqref{eq:env_coeff} depend on frequency, the quality factor will also depend on the tuning.
In order to quantify the influence of the voltage dependence of the
resonance frequency on $Q$, Fig.\ \ref{fig:Qtuning} shows
the quality factor for constant $\Omega_0=\Omega_0(0)$ (dashed lines) and
$\Omega_0(V_{\rm dc})$ (full lines). One sees that deviations between these two cases
appear only for larger voltages ($V_{\rm dc} > 20\;\text{V}$). Moreover, the qualitative
behavior and the cross-over from NLD to LD behavior remains unchanged. This confirms our
statement in Sec.\ \ref{sec:quality}, that the behavior of $Q$ is dominated by the damping
coefficients $\gamma$ and $\eta$ rather than the voltage dependence of $\Omega_0$.

Additionally, Fig.\ \ref{fig:Qtuning} shows the quality factor for a smaller value of the initial tension. In this case, the quality factor is decreased for all values of the static bias voltage. In the limit of large LD, this is due the increased static deflection (see Eq.\ \eqref{eq:q0lin}). In the opposite limit, the quality factor is independent of the static deflection, and the decrease in quality factor is instead a result of the decreasing resonance frequency $\Omega_0(0)\propto \sqrt{T_0}$. Furthermore, as argued at the end of Sec.\ \ref{sec:damping_ratio}, the cross-over between NLD and LD is shifted toward lower values of the bias voltage, signifying a decrease in the importance of NLD for lower tension.


\begin{thebibliography}{36}%
\makeatletter
\providecommand \@ifxundefined [1]{%
 \@ifx{#1\undefined}
}%
\providecommand \@ifnum [1]{%
 \ifnum #1\expandafter \@firstoftwo
 \else \expandafter \@secondoftwo
 \fi
}%
\providecommand \@ifx [1]{%
 \ifx #1\expandafter \@firstoftwo
 \else \expandafter \@secondoftwo
 \fi
}%
\providecommand \natexlab [1]{#1}%
\providecommand \enquote  [1]{``#1''}%
\providecommand \bibnamefont  [1]{#1}%
\providecommand \bibfnamefont [1]{#1}%
\providecommand \citenamefont [1]{#1}%
\providecommand \href@noop [0]{\@secondoftwo}%
\providecommand \href [0]{\begingroup \@sanitize@url \@href}%
\providecommand \@href[1]{\@@startlink{#1}\@@href}%
\providecommand \@@href[1]{\endgroup#1\@@endlink}%
\providecommand \@sanitize@url [0]{\catcode `\\12\catcode `\$12\catcode
  `\&12\catcode `\#12\catcode `\^12\catcode `\_12\catcode `\%12\relax}%
\providecommand \@@startlink[1]{}%
\providecommand \@@endlink[0]{}%
\providecommand \url  [0]{\begingroup\@sanitize@url \@url }%
\providecommand \@url [1]{\endgroup\@href {#1}{\urlprefix }}%
\providecommand \urlprefix  [0]{URL }%
\providecommand \Eprint [0]{\href }%
\providecommand \doibase [0]{http://dx.doi.org/}%
\providecommand \selectlanguage [0]{\@gobble}%
\providecommand \bibinfo  [0]{\@secondoftwo}%
\providecommand \bibfield  [0]{\@secondoftwo}%
\providecommand \translation [1]{[#1]}%
\providecommand \BibitemOpen [0]{}%
\providecommand \bibitemStop [0]{}%
\providecommand \bibitemNoStop [0]{.\EOS\space}%
\providecommand \EOS [0]{\spacefactor3000\relax}%
\providecommand \BibitemShut  [1]{\csname bibitem#1\endcsname}%
\let\auto@bib@innerbib\@empty
\bibitem [{\citenamefont {Bunch}\ \emph {et~al.}(2007)\citenamefont {Bunch},
  \citenamefont {van~der Zande}, \citenamefont {Verbridge}, \citenamefont
  {Frank}, \citenamefont {Tanenbaum}, \citenamefont {Parpia}, \citenamefont
  {Craighead},\ and\ \citenamefont {McEuen}}]{buza+07}%
  \BibitemOpen
  \bibfield  {author} {\bibinfo {author} {\bibfnamefont {J.~S.}\ \bibnamefont
  {Bunch}}, \bibinfo {author} {\bibfnamefont {A.~M.}\ \bibnamefont {van~der
  Zande}}, \bibinfo {author} {\bibfnamefont {S.~S.}\ \bibnamefont {Verbridge}},
  \bibinfo {author} {\bibfnamefont {I.~W.}\ \bibnamefont {Frank}}, \bibinfo
  {author} {\bibfnamefont {D.~M.}\ \bibnamefont {Tanenbaum}}, \bibinfo {author}
  {\bibfnamefont {J.~M.}\ \bibnamefont {Parpia}}, \bibinfo {author}
  {\bibfnamefont {H.~G.}\ \bibnamefont {Craighead}}, \ and\ \bibinfo {author}
  {\bibfnamefont {P.~L.}\ \bibnamefont {McEuen}},\ }\href {\doibase
  10.1126/science.1136836} {\bibfield  {journal} {\bibinfo  {journal}
  {Science}\ }\textbf {\bibinfo {volume} {315}},\ \bibinfo {pages} {490}
  (\bibinfo {year} {2007})}\BibitemShut {NoStop}%
\bibitem [{\citenamefont {Eriksson}\ \emph {et~al.}(2008)\citenamefont
  {Eriksson}, \citenamefont {Lee}, \citenamefont {Sourab}, \citenamefont
  {Isacsson}, \citenamefont {Kaunisto}, \citenamefont {Kinaret},\ and\
  \citenamefont {Campbell}}]{erle+08}%
  \BibitemOpen
  \bibfield  {author} {\bibinfo {author} {\bibfnamefont {A.}~\bibnamefont
  {Eriksson}}, \bibinfo {author} {\bibfnamefont {S.}~\bibnamefont {Lee}},
  \bibinfo {author} {\bibfnamefont {A.~A.}\ \bibnamefont {Sourab}}, \bibinfo
  {author} {\bibfnamefont {A.}~\bibnamefont {Isacsson}}, \bibinfo {author}
  {\bibfnamefont {R.}~\bibnamefont {Kaunisto}}, \bibinfo {author}
  {\bibfnamefont {J.~M.}\ \bibnamefont {Kinaret}}, \ and\ \bibinfo {author}
  {\bibfnamefont {E.~E.~B.}\ \bibnamefont {Campbell}},\ }\href {\doibase
  10.1021/nl080345w} {\bibfield  {journal} {\bibinfo  {journal} {Nano Lett.}\
  }\textbf {\bibinfo {volume} {8}},\ \bibinfo {pages} {1224} (\bibinfo {year}
  {2008})}\BibitemShut {NoStop}%
\bibitem [{\citenamefont {Chen}\ \emph {et~al.}(2009)\citenamefont {Chen},
  \citenamefont {Rosenblatt}, \citenamefont {Bolotin}, \citenamefont {Kalb},
  \citenamefont {Kim}, \citenamefont {Kymissis}, \citenamefont {Stormer},
  \citenamefont {Heinz},\ and\ \citenamefont {Hone}}]{chro+09}%
  \BibitemOpen
  \bibfield  {author} {\bibinfo {author} {\bibfnamefont {C.}~\bibnamefont
  {Chen}}, \bibinfo {author} {\bibfnamefont {S.}~\bibnamefont {Rosenblatt}},
  \bibinfo {author} {\bibfnamefont {K.~I.}\ \bibnamefont {Bolotin}}, \bibinfo
  {author} {\bibfnamefont {W.}~\bibnamefont {Kalb}}, \bibinfo {author}
  {\bibfnamefont {P.}~\bibnamefont {Kim}}, \bibinfo {author} {\bibfnamefont
  {I.}~\bibnamefont {Kymissis}}, \bibinfo {author} {\bibfnamefont {H.~L.}\
  \bibnamefont {Stormer}}, \bibinfo {author} {\bibfnamefont {T.~F.}\
  \bibnamefont {Heinz}}, \ and\ \bibinfo {author} {\bibfnamefont
  {J.}~\bibnamefont {Hone}},\ }\href@noop {} {\bibfield  {journal} {\bibinfo
  {journal} {Nat. Nanotechnol.}\ }\textbf {\bibinfo {volume} {4}},\ \bibinfo
  {pages} {861} (\bibinfo {year} {2009})}\BibitemShut {NoStop}%
\bibitem [{\citenamefont {Eichler}\ \emph {et~al.}(2011)\citenamefont
  {Eichler}, \citenamefont {Moser}, \citenamefont {Chaste}, \citenamefont
  {Zdrojek}, \citenamefont {Wilson-Rae},\ and\ \citenamefont
  {Bachtold}}]{eimo+11}%
  \BibitemOpen
  \bibfield  {author} {\bibinfo {author} {\bibfnamefont {A.}~\bibnamefont
  {Eichler}}, \bibinfo {author} {\bibfnamefont {J.}~\bibnamefont {Moser}},
  \bibinfo {author} {\bibfnamefont {J.}~\bibnamefont {Chaste}}, \bibinfo
  {author} {\bibfnamefont {M.}~\bibnamefont {Zdrojek}}, \bibinfo {author}
  {\bibfnamefont {I.}~\bibnamefont {Wilson-Rae}}, \ and\ \bibinfo {author}
  {\bibfnamefont {A.}~\bibnamefont {Bachtold}},\ }\href {\doibase
  10.1038/nnano.2011.71} {\bibfield  {journal} {\bibinfo  {journal} {Nat.
  Nanotechnol.}\ }\textbf {\bibinfo {volume} {6}},\ \bibinfo {pages} {339}
  (\bibinfo {year} {2011})}\BibitemShut {NoStop}%
\bibitem [{\citenamefont {Lifshitz}\ and\ \citenamefont
  {Roukes}(2000)}]{liro00}%
  \BibitemOpen
  \bibfield  {author} {\bibinfo {author} {\bibfnamefont {R.}~\bibnamefont
  {Lifshitz}}\ and\ \bibinfo {author} {\bibfnamefont {M.~L.}\ \bibnamefont
  {Roukes}},\ }\href {\doibase 10.1103/PhysRevB.61.5600} {\bibfield  {journal}
  {\bibinfo  {journal} {Phys. Rev. B}\ }\textbf {\bibinfo {volume} {61}},\
  \bibinfo {pages} {5600} (\bibinfo {year} {2000})}\BibitemShut {NoStop}%
\bibitem [{\citenamefont {Cross}\ and\ \citenamefont
  {Lifshitz}(2001)}]{crli01}%
  \BibitemOpen
  \bibfield  {author} {\bibinfo {author} {\bibfnamefont {M.~C.}\ \bibnamefont
  {Cross}}\ and\ \bibinfo {author} {\bibfnamefont {R.}~\bibnamefont
  {Lifshitz}},\ }\href {\doibase 10.1103/PhysRevB.64.085324} {\bibfield
  {journal} {\bibinfo  {journal} {Phys. Rev. B}\ }\textbf {\bibinfo {volume}
  {64}},\ \bibinfo {pages} {085324} (\bibinfo {year} {2001})}\BibitemShut
  {NoStop}%
\bibitem [{\citenamefont {Wilson-Rae}(2008)}]{wil08}%
  \BibitemOpen
  \bibfield  {author} {\bibinfo {author} {\bibfnamefont {I.}~\bibnamefont
  {Wilson-Rae}},\ }\href {\doibase 10.1103/PhysRevB.77.245418} {\bibfield
  {journal} {\bibinfo  {journal} {Phys. Rev. B}\ }\textbf {\bibinfo {volume}
  {77}},\ \bibinfo {pages} {245418} (\bibinfo {year} {2008})}\BibitemShut
  {NoStop}%
\bibitem [{\citenamefont {Remus}\ \emph {et~al.}(2009)\citenamefont {Remus},
  \citenamefont {Blencowe},\ and\ \citenamefont {Tanaka}}]{rebl+09}%
  \BibitemOpen
  \bibfield  {author} {\bibinfo {author} {\bibfnamefont {L.~G.}\ \bibnamefont
  {Remus}}, \bibinfo {author} {\bibfnamefont {M.~P.}\ \bibnamefont {Blencowe}},
  \ and\ \bibinfo {author} {\bibfnamefont {Y.}~\bibnamefont {Tanaka}},\ }\href
  {\doibase 10.1103/PhysRevB.80.174103} {\bibfield  {journal} {\bibinfo
  {journal} {Phys. Rev. B}\ }\textbf {\bibinfo {volume} {80}},\ \bibinfo
  {pages} {174103} (\bibinfo {year} {2009})}\BibitemShut {NoStop}%
\bibitem [{\citenamefont {Seo\'anez}\ \emph {et~al.}(2007)\citenamefont
  {Seo\'anez}, \citenamefont {Guinea},\ and\ \citenamefont
  {Castro~Neto}}]{segu+07}%
  \BibitemOpen
  \bibfield  {author} {\bibinfo {author} {\bibfnamefont {C.}~\bibnamefont
  {Seo\'anez}}, \bibinfo {author} {\bibfnamefont {F.}~\bibnamefont {Guinea}}, \
  and\ \bibinfo {author} {\bibfnamefont {A.~H.}\ \bibnamefont {Castro~Neto}},\
  }\href {\doibase 10.1103/PhysRevB.76.125427} {\bibfield  {journal} {\bibinfo
  {journal} {Phys. Rev. B}\ }\textbf {\bibinfo {volume} {76}},\ \bibinfo
  {pages} {125427} (\bibinfo {year} {2007})}\BibitemShut {NoStop}%
\bibitem [{\citenamefont {O'Connell}\ \emph {et~al.}(2010)\citenamefont
  {O'Connell}, \citenamefont {Hofheinz}, \citenamefont {Ansmann}, \citenamefont
  {Bialczak}, \citenamefont {Lenander}, \citenamefont {Lucero}, \citenamefont
  {Neeley}, \citenamefont {Sank}, \citenamefont {Wang}, \citenamefont {Weides},
  \citenamefont {Wenner}, \citenamefont {Martinis},\ and\ \citenamefont
  {Cleland}}]{ocho+10}%
  \BibitemOpen
  \bibfield  {author} {\bibinfo {author} {\bibfnamefont {A.~D.}\ \bibnamefont
  {O'Connell}}, \bibinfo {author} {\bibfnamefont {M.}~\bibnamefont {Hofheinz}},
  \bibinfo {author} {\bibfnamefont {M.}~\bibnamefont {Ansmann}}, \bibinfo
  {author} {\bibfnamefont {R.~C.}\ \bibnamefont {Bialczak}}, \bibinfo {author}
  {\bibfnamefont {M.}~\bibnamefont {Lenander}}, \bibinfo {author}
  {\bibfnamefont {E.}~\bibnamefont {Lucero}}, \bibinfo {author} {\bibfnamefont
  {M.}~\bibnamefont {Neeley}}, \bibinfo {author} {\bibfnamefont
  {D.}~\bibnamefont {Sank}}, \bibinfo {author} {\bibfnamefont {H.}~\bibnamefont
  {Wang}}, \bibinfo {author} {\bibfnamefont {M.}~\bibnamefont {Weides}},
  \bibinfo {author} {\bibfnamefont {J.}~\bibnamefont {Wenner}}, \bibinfo
  {author} {\bibfnamefont {J.~M.}\ \bibnamefont {Martinis}}, \ and\ \bibinfo
  {author} {\bibfnamefont {A.~N.}\ \bibnamefont {Cleland}},\ }\href {\doibase
  10.1038/nature08967} {\bibfield  {journal} {\bibinfo  {journal} {Nature}\
  }\textbf {\bibinfo {volume} {464}},\ \bibinfo {pages} {697} (\bibinfo {year}
  {2010})}\BibitemShut {NoStop}%
\bibitem [{\citenamefont {Teufel}\ \emph {et~al.}(2011)\citenamefont {Teufel},
  \citenamefont {Donner}, \citenamefont {Li}, \citenamefont {Harlow},
  \citenamefont {Allman}, \citenamefont {Cicak}, \citenamefont {Sirois},
  \citenamefont {Whittaker}, \citenamefont {Lehnert},\ and\ \citenamefont
  {Simmonds}}]{tedo+11}%
  \BibitemOpen
  \bibfield  {author} {\bibinfo {author} {\bibfnamefont {J.~D.}\ \bibnamefont
  {Teufel}}, \bibinfo {author} {\bibfnamefont {T.}~\bibnamefont {Donner}},
  \bibinfo {author} {\bibfnamefont {D.}~\bibnamefont {Li}}, \bibinfo {author}
  {\bibfnamefont {J.~W.}\ \bibnamefont {Harlow}}, \bibinfo {author}
  {\bibfnamefont {M.~S.}\ \bibnamefont {Allman}}, \bibinfo {author}
  {\bibfnamefont {K.}~\bibnamefont {Cicak}}, \bibinfo {author} {\bibfnamefont
  {A.~J.}\ \bibnamefont {Sirois}}, \bibinfo {author} {\bibfnamefont {J.~D.}\
  \bibnamefont {Whittaker}}, \bibinfo {author} {\bibfnamefont {K.~W.}\
  \bibnamefont {Lehnert}}, \ and\ \bibinfo {author} {\bibfnamefont {R.~W.}\
  \bibnamefont {Simmonds}},\ }\href {\doibase 10.1038/nature10261} {\bibfield
  {journal} {\bibinfo  {journal} {Nature}\ }\textbf {\bibinfo {volume} {475}},\
  \bibinfo {pages} {359} (\bibinfo {year} {2011})}\BibitemShut {NoStop}%
\bibitem [{\citenamefont {Voje}\ \emph {et~al.}(2012)\citenamefont {Voje},
  \citenamefont {Kinaret},\ and\ \citenamefont {Isacsson}}]{voki+12}%
  \BibitemOpen
  \bibfield  {author} {\bibinfo {author} {\bibfnamefont {A.}~\bibnamefont
  {Voje}}, \bibinfo {author} {\bibfnamefont {J.~M.}\ \bibnamefont {Kinaret}}, \
  and\ \bibinfo {author} {\bibfnamefont {A.}~\bibnamefont {Isacsson}},\
  }\href@noop {} {\bibfield  {journal} {\bibinfo  {journal} {Phys. Rev. B}\
  }\textbf {\bibinfo {volume} {85}},\ \bibinfo {pages} {205415} (\bibinfo
  {year} {2012})}\BibitemShut {NoStop}%
\bibitem [{\citenamefont {Dykman}\ and\ \citenamefont
  {Krivoglaz}(1984)}]{dykr84}%
  \BibitemOpen
  \bibfield  {author} {\bibinfo {author} {\bibfnamefont {M.}~\bibnamefont
  {Dykman}}\ and\ \bibinfo {author} {\bibfnamefont {M.}~\bibnamefont
  {Krivoglaz}},\ }\href@noop {} {\bibfield  {journal} {\bibinfo  {journal}
  {Soviet Scientific Reviews, Section A, Physics Reviews}\ }\textbf {\bibinfo
  {volume} {5}},\ \bibinfo {pages} {265} (\bibinfo {year} {1984})}\BibitemShut
  {NoStop}%
\bibitem [{\citenamefont {Lifshitz}\ and\ \citenamefont
  {Cross}(2008)}]{licr08}%
  \BibitemOpen
  \bibfield  {author} {\bibinfo {author} {\bibfnamefont {R.}~\bibnamefont
  {Lifshitz}}\ and\ \bibinfo {author} {\bibfnamefont {M.}~\bibnamefont
  {Cross}},\ }\enquote {\bibinfo {title} {Nonlinear dynamics of nanomechanical
  and micromechanical resonators},}\ \ (\bibinfo  {publisher} {Wiley-VCH},\
  \bibinfo {year} {2008})\ Chap.~\bibinfo {chapter} {1}\BibitemShut {NoStop}%
\bibitem [{\citenamefont {Zaitsev}\ \emph {et~al.}(2012)\citenamefont
  {Zaitsev}, \citenamefont {Shtempluck}, \citenamefont {Buks},\ and\
  \citenamefont {Gottlieb}}]{zash+12}%
  \BibitemOpen
  \bibfield  {author} {\bibinfo {author} {\bibfnamefont {S.}~\bibnamefont
  {Zaitsev}}, \bibinfo {author} {\bibfnamefont {O.}~\bibnamefont {Shtempluck}},
  \bibinfo {author} {\bibfnamefont {E.}~\bibnamefont {Buks}}, \ and\ \bibinfo
  {author} {\bibfnamefont {O.}~\bibnamefont {Gottlieb}},\ }\href@noop {}
  {\bibfield  {journal} {\bibinfo  {journal} {Nonlinear Dynam.}\ }\textbf
  {\bibinfo {volume} {67}},\ \bibinfo {pages} {859} (\bibinfo {year}
  {2012})}\BibitemShut {NoStop}%
\bibitem [{\citenamefont {Zwanzig}(1973)}]{zw73}%
  \BibitemOpen
  \bibfield  {author} {\bibinfo {author} {\bibfnamefont {R.}~\bibnamefont
  {Zwanzig}},\ }\href {\doibase 10.1007/BF01008729} {\bibfield  {journal}
  {\bibinfo  {journal} {J. Stat. Phys.}\ }\textbf {\bibinfo {volume} {9}},\
  \bibinfo {pages} {215} (\bibinfo {year} {1973})}\BibitemShut {NoStop}%
\bibitem [{\citenamefont {Lindenberg}\ and\ \citenamefont
  {Seshadri}(1981)}]{lise81}%
  \BibitemOpen
  \bibfield  {author} {\bibinfo {author} {\bibfnamefont {K.}~\bibnamefont
  {Lindenberg}}\ and\ \bibinfo {author} {\bibfnamefont {V.}~\bibnamefont
  {Seshadri}},\ }\href {\doibase 10.1016/0378-4371(81)90007-8} {\bibfield
  {journal} {\bibinfo  {journal} {Physica A}\ }\textbf {\bibinfo {volume}
  {109}},\ \bibinfo {pages} {483} (\bibinfo {year} {1981})}\BibitemShut
  {NoStop}%
\bibitem [{\citenamefont {Castro~Neto}\ \emph {et~al.}(2009)\citenamefont
  {Castro~Neto}, \citenamefont {Guinea}, \citenamefont {Peres}, \citenamefont
  {Novoselov},\ and\ \citenamefont {Geim}}]{cagu+09}%
  \BibitemOpen
  \bibfield  {author} {\bibinfo {author} {\bibfnamefont {A.~H.}\ \bibnamefont
  {Castro~Neto}}, \bibinfo {author} {\bibfnamefont {F.}~\bibnamefont {Guinea}},
  \bibinfo {author} {\bibfnamefont {N.~M.~R.}\ \bibnamefont {Peres}}, \bibinfo
  {author} {\bibfnamefont {K.~S.}\ \bibnamefont {Novoselov}}, \ and\ \bibinfo
  {author} {\bibfnamefont {A.~K.}\ \bibnamefont {Geim}},\ }\href {\doibase
  10.1103/RevModPhys.81.109} {\bibfield  {journal} {\bibinfo  {journal} {Rev.
  Mod. Phys.}\ }\textbf {\bibinfo {volume} {81}},\ \bibinfo {pages} {109}
  (\bibinfo {year} {2009})}\BibitemShut {NoStop}%
\bibitem [{\citenamefont {von Oppen}\ \emph {et~al.}(2009)\citenamefont {von
  Oppen}, \citenamefont {Guinea},\ and\ \citenamefont {Mariani}}]{vogu+09}%
  \BibitemOpen
  \bibfield  {author} {\bibinfo {author} {\bibfnamefont {F.}~\bibnamefont {von
  Oppen}}, \bibinfo {author} {\bibfnamefont {F.}~\bibnamefont {Guinea}}, \ and\
  \bibinfo {author} {\bibfnamefont {E.}~\bibnamefont {Mariani}},\ }\href
  {\doibase 10.1103/PhysRevB.80.075420} {\bibfield  {journal} {\bibinfo
  {journal} {Phys. Rev. B}\ }\textbf {\bibinfo {volume} {80}},\ \bibinfo
  {pages} {075420} (\bibinfo {year} {2009})}\BibitemShut {NoStop}%
\bibitem [{\citenamefont {Sabio}\ \emph {et~al.}(2008)\citenamefont {Sabio},
  \citenamefont {Seo\'anez}, \citenamefont {Fratini}, \citenamefont {Guinea},
  \citenamefont {Castro~Neto},\ and\ \citenamefont {Sols}}]{sase+08}%
  \BibitemOpen
  \bibfield  {author} {\bibinfo {author} {\bibfnamefont {J.}~\bibnamefont
  {Sabio}}, \bibinfo {author} {\bibfnamefont {C.}~\bibnamefont {Seo\'anez}},
  \bibinfo {author} {\bibfnamefont {S.}~\bibnamefont {Fratini}}, \bibinfo
  {author} {\bibfnamefont {F.}~\bibnamefont {Guinea}}, \bibinfo {author}
  {\bibfnamefont {A.~H.}\ \bibnamefont {Castro~Neto}}, \ and\ \bibinfo {author}
  {\bibfnamefont {F.}~\bibnamefont {Sols}},\ }\href {\doibase
  10.1103/PhysRevB.77.195409} {\bibfield  {journal} {\bibinfo  {journal} {Phys.
  Rev. B}\ }\textbf {\bibinfo {volume} {77}},\ \bibinfo {pages} {195409}
  (\bibinfo {year} {2008})}\BibitemShut {NoStop}%
\bibitem [{\citenamefont {Koenig}\ \emph {et~al.}(2011)\citenamefont {Koenig},
  \citenamefont {Boddeti}, \citenamefont {Dunn},\ and\ \citenamefont
  {Bunch}}]{kobo+11}%
  \BibitemOpen
  \bibfield  {author} {\bibinfo {author} {\bibfnamefont {S.~P.}\ \bibnamefont
  {Koenig}}, \bibinfo {author} {\bibfnamefont {N.~G.}\ \bibnamefont {Boddeti}},
  \bibinfo {author} {\bibfnamefont {M.~L.}\ \bibnamefont {Dunn}}, \ and\
  \bibinfo {author} {\bibfnamefont {J.~S.}\ \bibnamefont {Bunch}},\ }\href
  {\doibase 10.1038/nnano.2011.123} {\bibfield  {journal} {\bibinfo  {journal}
  {Nat. Nanotechnol.}\ }\textbf {\bibinfo {volume} {6}},\ \bibinfo {pages}
  {543} (\bibinfo {year} {2011})}\BibitemShut {NoStop}%
\bibitem [{\citenamefont {Viola~Kusminskiy}\ \emph {et~al.}(2011)\citenamefont
  {Viola~Kusminskiy}, \citenamefont {Campbell}, \citenamefont {Castro~Neto},\
  and\ \citenamefont {Guinea}}]{kuca+11}%
  \BibitemOpen
  \bibfield  {author} {\bibinfo {author} {\bibfnamefont {S.}~\bibnamefont
  {Viola~Kusminskiy}}, \bibinfo {author} {\bibfnamefont {D.~K.}\ \bibnamefont
  {Campbell}}, \bibinfo {author} {\bibfnamefont {A.~H.}\ \bibnamefont
  {Castro~Neto}}, \ and\ \bibinfo {author} {\bibfnamefont {F.}~\bibnamefont
  {Guinea}},\ }\href {\doibase 10.1103/PhysRevB.83.165405} {\bibfield
  {journal} {\bibinfo  {journal} {Phys. Rev. B}\ }\textbf {\bibinfo {volume}
  {83}},\ \bibinfo {pages} {165405} (\bibinfo {year} {2011})}\BibitemShut
  {NoStop}%
\bibitem [{\citenamefont {Nelson}\ \emph {et~al.}(1989)\citenamefont {Nelson},
  \citenamefont {Piran},\ and\ \citenamefont {Weinberg}}]{nepi+88}%
  \BibitemOpen
  \bibinfo {editor} {\bibfnamefont {D.}~\bibnamefont {Nelson}}, \bibinfo
  {editor} {\bibfnamefont {T.}~\bibnamefont {Piran}}, \ and\ \bibinfo {editor}
  {\bibfnamefont {S.}~\bibnamefont {Weinberg}},\ eds.,\ \href@noop {} {\emph
  {\bibinfo {title} {Statistical Mechanics of Membranes and Surfaces}}}\
  (\bibinfo  {publisher} {World Scientific},\ \bibinfo {year}
  {1989})\BibitemShut {NoStop}%
\bibitem [{\citenamefont {Landau}\ and\ \citenamefont
  {Lifshitz}(1986)}]{lali86}%
  \BibitemOpen
  \bibfield  {author} {\bibinfo {author} {\bibfnamefont {L.~D.}\ \bibnamefont
  {Landau}}\ and\ \bibinfo {author} {\bibfnamefont {E.~M.}\ \bibnamefont
  {Lifshitz}},\ }\href@noop {} {\emph {\bibinfo {title} {Theory of
  elasticity}}},\ \bibinfo {edition} {3rd}\ ed.,\ edited by\ \bibinfo {editor}
  {\bibfnamefont {A.~M.}\ \bibnamefont {Kosevich}}\ and\ \bibinfo {editor}
  {\bibfnamefont {L.~P.}\ \bibnamefont {Pitaevski{\u\i}}}\ (\bibinfo
  {publisher} {Butterworth-Heinemann},\ \bibinfo {address} {Oxford},\ \bibinfo
  {year} {1986})\BibitemShut {NoStop}%
\bibitem [{\citenamefont {Yakobson}\ \emph {et~al.}(1996)\citenamefont
  {Yakobson}, \citenamefont {Brabec},\ and\ \citenamefont
  {Bernholc}}]{yabr+96}%
  \BibitemOpen
  \bibfield  {author} {\bibinfo {author} {\bibfnamefont {B.~I.}\ \bibnamefont
  {Yakobson}}, \bibinfo {author} {\bibfnamefont {C.~J.}\ \bibnamefont
  {Brabec}}, \ and\ \bibinfo {author} {\bibfnamefont {J.}~\bibnamefont
  {Bernholc}},\ }\href {\doibase 10.1103/PhysRevLett.76.2511} {\bibfield
  {journal} {\bibinfo  {journal} {Phys. Rev. Lett.}\ }\textbf {\bibinfo
  {volume} {76}},\ \bibinfo {pages} {2511} (\bibinfo {year}
  {1996})}\BibitemShut {NoStop}%
\bibitem [{\citenamefont {Fasolino}\ \emph {et~al.}(2007)\citenamefont
  {Fasolino}, \citenamefont {Los},\ and\ \citenamefont {Katsnelson}}]{falo+07}%
  \BibitemOpen
  \bibfield  {author} {\bibinfo {author} {\bibfnamefont {A.}~\bibnamefont
  {Fasolino}}, \bibinfo {author} {\bibfnamefont {J.~H.}\ \bibnamefont {Los}}, \
  and\ \bibinfo {author} {\bibfnamefont {M.~I.}\ \bibnamefont {Katsnelson}},\
  }\href {\doibase 10.1038/nmat2011} {\bibfield  {journal} {\bibinfo  {journal}
  {Nat. Mater.}\ }\textbf {\bibinfo {volume} {6}},\ \bibinfo {pages} {858}
  (\bibinfo {year} {2007})}\BibitemShut {NoStop}%
\bibitem [{\citenamefont {Atalaya}\ \emph {et~al.}(2008)\citenamefont
  {Atalaya}, \citenamefont {Isacsson},\ and\ \citenamefont
  {Kinaret}}]{atis+08}%
  \BibitemOpen
  \bibfield  {author} {\bibinfo {author} {\bibfnamefont {J.}~\bibnamefont
  {Atalaya}}, \bibinfo {author} {\bibfnamefont {A.}~\bibnamefont {Isacsson}}, \
  and\ \bibinfo {author} {\bibfnamefont {J.~M.}\ \bibnamefont {Kinaret}},\
  }\href {\doibase 10.1021/nl801733d} {\bibfield  {journal} {\bibinfo
  {journal} {Nano Lett.}\ }\textbf {\bibinfo {volume} {8}},\ \bibinfo {pages}
  {4196} (\bibinfo {year} {2008})}\BibitemShut {NoStop}%
\bibitem [{\citenamefont {Lindahl}\ \emph {et~al.}(2012)\citenamefont
  {Lindahl}, \citenamefont {Midtvedt}, \citenamefont {Svensson}, \citenamefont
  {Nerushev}, \citenamefont {Lindvall}, \citenamefont {Isacsson},\ and\
  \citenamefont {Campbell}}]{limi+12}%
  \BibitemOpen
  \bibfield  {author} {\bibinfo {author} {\bibfnamefont {N.}~\bibnamefont
  {Lindahl}}, \bibinfo {author} {\bibfnamefont {D.}~\bibnamefont {Midtvedt}},
  \bibinfo {author} {\bibfnamefont {J.}~\bibnamefont {Svensson}}, \bibinfo
  {author} {\bibfnamefont {O.~A.}\ \bibnamefont {Nerushev}}, \bibinfo {author}
  {\bibfnamefont {N.}~\bibnamefont {Lindvall}}, \bibinfo {author}
  {\bibfnamefont {A.}~\bibnamefont {Isacsson}}, \ and\ \bibinfo {author}
  {\bibfnamefont {E.~E.~B.}\ \bibnamefont {Campbell}},\ }\href {\doibase
  10.1021/nl301080v} {\bibfield  {journal} {\bibinfo  {journal} {Nano Lett.}\
  }\textbf {\bibinfo {volume} {12}},\ \bibinfo {pages} {3526} (\bibinfo {year}
  {2012})}\BibitemShut {NoStop}%
\bibitem [{\citenamefont {Swain}\ and\ \citenamefont
  {Andelman}(1999)}]{swan99}%
  \BibitemOpen
  \bibfield  {author} {\bibinfo {author} {\bibfnamefont {P.~S.}\ \bibnamefont
  {Swain}}\ and\ \bibinfo {author} {\bibfnamefont {D.}~\bibnamefont
  {Andelman}},\ }\href {\doibase 10.1021/la990503m} {\bibfield  {journal}
  {\bibinfo  {journal} {Langmuir}\ }\textbf {\bibinfo {volume} {15}},\ \bibinfo
  {pages} {8902} (\bibinfo {year} {1999})}\BibitemShut {NoStop}%
\bibitem [{\citenamefont {Persson}(2001)}]{pe01}%
  \BibitemOpen
  \bibfield  {author} {\bibinfo {author} {\bibfnamefont {B.~N.~J.}\
  \bibnamefont {Persson}},\ }\href {\doibase 10.1063/1.1388626} {\bibfield
  {journal} {\bibinfo  {journal} {J. Chem. Phys.}\ }\textbf {\bibinfo {volume}
  {115}},\ \bibinfo {pages} {3840} (\bibinfo {year} {2001})}\BibitemShut
  {NoStop}%
\bibitem [{\citenamefont {Maradudin}\ and\ \citenamefont
  {Mills}(1976)}]{mami76}%
  \BibitemOpen
  \bibfield  {author} {\bibinfo {author} {\bibfnamefont {A.}~\bibnamefont
  {Maradudin}}\ and\ \bibinfo {author} {\bibfnamefont {D.}~\bibnamefont
  {Mills}},\ }\href {\doibase 10.1016/0003-4916(76)90063-4} {\bibfield
  {journal} {\bibinfo  {journal} {Ann. Phys. - New York}\ }\textbf {\bibinfo
  {volume} {100}},\ \bibinfo {pages} {262 } (\bibinfo {year}
  {1976})}\BibitemShut {NoStop}%
\bibitem [{\citenamefont {Barnard}\ \emph {et~al.}(2012)\citenamefont
  {Barnard}, \citenamefont {Sazonova}, \citenamefont {van~der Zande},\ and\
  \citenamefont {McEuen}}]{basa+12}%
  \BibitemOpen
  \bibfield  {author} {\bibinfo {author} {\bibfnamefont {A.~W.}\ \bibnamefont
  {Barnard}}, \bibinfo {author} {\bibfnamefont {V.}~\bibnamefont {Sazonova}},
  \bibinfo {author} {\bibfnamefont {A.~M.}\ \bibnamefont {van~der Zande}}, \
  and\ \bibinfo {author} {\bibfnamefont {P.~L.}\ \bibnamefont {McEuen}},\
  }\href {\doibase 10.1073/pnas.1216407109} {\bibfield  {journal} {\bibinfo
  {journal} {PNAS}\ }\textbf {\bibinfo {volume} {109}},\ \bibinfo {pages}
  {19093} (\bibinfo {year} {2012})}\BibitemShut {NoStop}%
\bibitem [{Note1()}]{Note1}%
  \BibitemOpen
  \bibinfo {note} {We found that $\protect \overline {R}_{xx}(x-x',\omega )$ is
  well approximated by the integral $\DOTSI \intop \ilimits@ ^{b/2}_{-b/2}
  dy\protect \tmspace +\thinmuskip {.1667em} R_{xx}(x-x',y,\omega
  )$.}\BibitemShut {Stop}%
\bibitem [{\citenamefont {Press}\ \emph {et~al.}(1992)\citenamefont {Press},
  \citenamefont {Flannery}, \citenamefont {Teukolsky},\ and\ \citenamefont
  {Vetterling}}]{prfl+92}%
  \BibitemOpen
  \bibfield  {author} {\bibinfo {author} {\bibfnamefont {W.~H.}\ \bibnamefont
  {Press}}, \bibinfo {author} {\bibfnamefont {B.~P.}\ \bibnamefont {Flannery}},
  \bibinfo {author} {\bibfnamefont {S.~A.}\ \bibnamefont {Teukolsky}}, \ and\
  \bibinfo {author} {\bibfnamefont {W.~T.}\ \bibnamefont {Vetterling}},\
  }\href@noop {} {\emph {\bibinfo {title} {Numerical Recipes in {C}: {T}he Art
  of Scientific Computing}}}\ (\bibinfo  {publisher} {Cambridge University
  Press},\ \bibinfo {address} {Cambridge},\ \bibinfo {year} {1992})\ p.\
  \bibinfo {pages} {994}\BibitemShut {NoStop}%
\bibitem [{\citenamefont {Lee}\ \emph {et~al.}(2008)\citenamefont {Lee},
  \citenamefont {Wei}, \citenamefont {Kysar},\ and\ \citenamefont
  {Hone}}]{lewe+08}%
  \BibitemOpen
  \bibfield  {author} {\bibinfo {author} {\bibfnamefont {C.}~\bibnamefont
  {Lee}}, \bibinfo {author} {\bibfnamefont {X.}~\bibnamefont {Wei}}, \bibinfo
  {author} {\bibfnamefont {J.~W.}\ \bibnamefont {Kysar}}, \ and\ \bibinfo
  {author} {\bibfnamefont {J.}~\bibnamefont {Hone}},\ }\href {\doibase
  10.1126/science.1157996} {\bibfield  {journal} {\bibinfo  {journal}
  {Science}\ }\textbf {\bibinfo {volume} {321}},\ \bibinfo {pages} {385}
  (\bibinfo {year} {2008})}\BibitemShut {NoStop}%
\bibitem [{\citenamefont {Persson}\ and\ \citenamefont {Ueba}(2010)}]{peue10}%
  \BibitemOpen
  \bibfield  {author} {\bibinfo {author} {\bibfnamefont {B.~N.~J.}\
  \bibnamefont {Persson}}\ and\ \bibinfo {author} {\bibfnamefont
  {H.}~\bibnamefont {Ueba}},\ }\href
  {http://stacks.iop.org/0295-5075/91/i=5/a=56001} {\bibfield  {journal}
  {\bibinfo  {journal} {Europhys. Lett.}\ }\textbf {\bibinfo {volume} {91}},\
  \bibinfo {pages} {56001} (\bibinfo {year} {2010})}\BibitemShut {NoStop}%
\end{thebibliography}
%

\end{document}